\documentclass{article}
\usepackage{amsmath,amsfonts,amssymb}
\usepackage{array}
\usepackage{booktabs}
\usepackage{longtable}
\usepackage{calc}
\usepackage{graphicx}
\usepackage{textcomp}
\usepackage{url}
\usepackage{hyperref}
\usepackage{multirow}
\usepackage{multicol}
\usepackage{algorithm2e}
\usepackage{adjustbox}
\usepackage{tabularx}
\usepackage{caption}
\usepackage{authblk}
\usepackage{array}
\usepackage{geometry}
\hypersetup{hidelinks}
\usepackage[table]{xcolor}
\definecolor {processblue}{cmyk}{0.60,0.55,0.60,0}

\usepackage{indentfirst,mathrsfs,amscd}

\usepackage{tikz-cd}
\usepackage{circuitikz}
\usetikzlibrary{shapes.geometric, arrows}

\newtheorem{ex}{Example}{}

\title{Theory Building in Software Engineering: Operationalization}
\author[1]{Jorge P\'erez}
\author[1]{Jessica D\'iaz}
\author[2,3]{\'Angel Gonz\'alez-Prieto}
\author[1]{Sergio Gil-Borr\'as}
\affil[1]{ETSI Sistemas Inform{\'a}ticos, Universidad Politécnica de Madrid.}
\affil[2]{Facultad de Ciencias Matem\'aticas\\
    Universidad Complutense de Madrid.}
\affil[3]{Instituto de Ciencias Matem\'aticas (CSIC-UAM-UCM-UC3M).}

\date{}

\begin{document}
\maketitle

\begin{abstract}
This work is part of a research project whose ultimate goal is to systematize a procedure for constructing theories in software engineering. The proposed methodology involves four phases: conceptualization, operationalization, testing, and application. In previous work, we described the conceptualization process. This paper presents a set of procedures for systematizing the operationalization phase in theory building. Specifically, it translates the concepts and propositions obtained from the previous conceptualization into constructs and empirically testable hypotheses. The operationalization phase is structured across three distinct dimensions: the practical procedural steps, their strict formal mathematical specification, and their evaluation through an illustrative empirical case study. Grounded in critical realism, we operationalize the qualitatively derived concepts and propositions into explicit constructs and logical hypotheses. It is important to emphasize that a novel strategy has been designed that reduces the number of hypotheses resulting from this process. In the use case described, more than 30\% of the initial hypotheses have been eliminated because they are redundant. This paper is a pioneering contribution in offering comprehensive guidelines for theory operationalization using logical implication and describes a strategy based on graph theory that tackles the problem of hypothesis explosion and maintains the principle of parsimony.\\[0.15cm]
\end{abstract}

\begin{multicols}{2}

\section{Introduction}\label{introduction}

Theories are essential because they provide structured explanations of phenomena, helping researchers make sense of complex realities. \emph{``Theory lies at the heart of a scholarly discipline, supporting its scholarly relevance, identity and legitimacy. Without theory and the associated cumulative contribution to knowledge, the viability of a discipline is jeopardized because its scholarly distinctiveness is lost''} \cite{diaz2023importance}.

Despite the importance of theories, there is no consensus on either their definition or on how they should be constructed. What exists are micro-theories, such as those in the subfield of thermodynamics (within physics), or those related to recurring methods and activities in software engineering projects, such as formalized micro-theories in requirements engineering, program verification, software testing, and similar areas. The existence of micro-theories for smaller subfields highlights that their construction is indeed possible. Nevertheless, there appears to be no agreed-upon guide on how to develop these micro-theories (hereafter referred to simply as theory), nor are there established guidelines on how to express them.

The research program to which this work belongs aims to contribute to the field of software engineering by specifying a method for constructing, validating, and applying theories. This paper focuses on describing a method to support one of the essential phases in theory construction: operationalization. Our approach to operationalizing concepts and propositions presents two novelties: i) the use of logical implication instead of causal reasoning (see Appendix \ref{app:a1-main}); and ii) a strategy, based on graph operations, that automatically reduces the final number of hypotheses, thereby preserving the principle of parsimony. Both novelties are intrinsically related. Hypotheses expressed in the form of logical implication allow all of them to be represented with a directed graph; and there are operations on graphs to obtain the minimum graph equivalent to the original graph.

This paper is structured as follows. Section \ref{background-and-related-work} presents a background for building theories and an overview of the approaches to the operationalization problem. Section \ref{sec:framework} describes our proposed methodological framework for theory building. Section \ref{research-methodology} describes and justifies the methodology employed in this investigation. In Section \ref{operationalization-1} we address the operationalization of both concepts and propositions of a theory, illustrated with a simple case study. In Section \ref{results}, we discuss our main results. The limitations and threats to the validity of this research, together with the strategies implemented to mitigate them, are described in Section \ref{threats-to-validity-reliability-and-limitations}. Finally, Section \ref{conclusion-future-works} presents our main conclusions from this research and further work.

\section{Background and Related Work}\label{background-and-related-work}

This section first presents several definitions of the term theory, along with its constituent elements and a typology of theories. We then describe the theory-building process, with particular emphasis on the operationalization phase.

\subsection{Theory: elements and types}\label{theory-elements-and-types}

The term theory is polysemic, meaning different things to different stakeholders. A general and universal definition is offered by Metcalfe \cite[p.~13]{n4}: ``\emph{Interpretations of what is theory include: `an explanation why', a hypothesis or a law to a scientist, `ideally or hopefully' to a practitioner, `a knowledge claim or an argument' to an epistemologist, `an ideology' to a Marxist, a `perspective' to a systems thinker, `the interrelationship between the notation of music and performance practice' to a musician, `a verb not a noun' to Karl Weick, `policy' to a bureaucrat, and `an explanation how to structure' to a designer}.'' And just as there are multiple definitions of theory (Appendix \ref{app:a1-theory}), different authors propose various elements that contribute to its composition (Appendix \ref{app:a2}) and establish different taxonomies of theories (Appendix \ref{app:a2-types}).

For this research, we define a theory as a set of constructs related through hypotheses within a defined scope or boundary. All relationships must be empirically testable and falsifiable, in line with Popper. We adopt Schöning's \cite{n9} formalism to express the theory. For this operationalization phase, we adopt a critical realist stance, utilizing logical formalization not as a tool for positivist metric quantification, but as an analytical scaffolding to map and optimize the structural dependencies of generative mechanisms in open systems.

\subsection{Theory building process}\label{theory-building-process}

This section describes some models for theory construction: what is known as ``theorizing''. Theorizing \emph{``refers to the evolution of theory over time as well as to theory emergence. Product (theory) and process (theorizing) do not have to be competing or contradicting views; they are complementary''} \cite{mueller2017understanding}.

In this sense, theoretical reasoning (theorizing) can be conceived as the process of using existing knowledge to draw conclusions, make predictions, or construct explanations, which aligns with the different types of theories indicated in the Appendix \ref{app:a2-types}. This reasoning can be inductive, deductive or abductive.

With regard to the products generated during the theorizing process (i.e., theory building), it is worth mentioning the work of Hassan, Lowry \& Mathiassen \cite{hassan2022useful}, which, following Weick \cite{weick1995theory}, prioritizes intermediate (or pretheoretical) products over theory as an end product. They identify 12 such products: question, paradigm, law, framework, myth, analogy, metaphor, model, concept, construct, statement, and hypothesis.

Appendix \ref{theory-building-process-1} describes three proposals for developing theories: Lynham's general method of theory construction \cite{lynham2002} (Section \ref{sec:lynham-general}), Ralph's process theories \cite{ralph:2019} (Section \ref{sec:ralph-process}) and Rivard's spiral model \cite{rivard2021theory} (Section \ref{sec:rivard-spiral}). While these landmark proposals by Lynham, Ralph, and Rivard provide robust foundations for general theory building, they present a significant research gap regarding the mechanics of the operationalization phase. Lynham identifies operationalization as a distinct phase but offers little procedural guidance on how to execute it. Ralph focuses heavily on qualitative data analysis, while Rivard emphasizes causal explanations without providing tools to manage the complexity of those relationships. None of these existing frameworks provide a systematic method to transition from qualitative propositions to formal hypotheses while addressing the cognitive overload caused by hypothesis explosion in open empirical systems. Our proposal directly addresses this gap by introducing a critical realist, graph-theoretic approach that formalizes this transition while automatically preserving theoretical parsimony through logical reduction

\subsection{Operationalization}\label{operationalization}

During the operationalization process, it is important to distinguish between the following terms: concept, variables, construct, proposition, and hypothesis. A construct is a concept that has been operationalized into variables \cite{sjoberg2023}. A hypothesis represents a prediction about the values that the variables associated with the concepts of a theory will take; it uses empirical values for the variables mentioned in each proposition \cite{dubin:1978}. In other words, a construct is a concept with associated variables and a defined set of possible values (empirical indicators). A hypothesis expresses a relationship between constructs (i.e., the relationship between concepts expressed in the proposition) such that the values of the variables of each construct are related.

The purpose of the operationalization phase of theory-building research is to make an explicit connection between the conceptualization phase and practice: the theoretical framework must be translated or converted into observable and confirmable components \cite[p. 232]{lynham2002}. Thus, operationalization addresses two aspects \cite{sjoberg:2008}:

\begin{itemize}
\item
  Operationalizing theoretical concepts into constructs.
\item
  Operationalizing theoretical propositions into empirically testable hypotheses.
\end{itemize}

\subsubsection{Operationalizing theoretical concepts} According to Sjøberg and Bergersen \cite{sjoberg2023}, in empirical science it is often necessary to measure concepts. Although some concepts are directly measurable, others require the use of indicators (variables) to represent them at the observational level. ``The process of determining such indicators is called operationalization. A construct is a concept that has an associated operationalization into indicators'' \cite[p. 1375]{sjoberg2023}.

Operationalizing the concepts involves three steps \cite{sjoberg2023}: (i) identify the concepts to operationalize; (ii) choose one or more variables to represent each concept; and (iii) select values for each variable (i.e., define the range of values or type they can take at different times).

\subsubsection{Operationalizing theoretical propositions} Operationalizing propositions involves converting them into testable hypotheses by identifying and linking the relevant variables and values that were previously defined. Recently, Kanaparan and Strode \cite{Kanaparan:2021} outlined some guidelines to effectively convert propositions into hypotheses and described a structured approach to ensure the integrity and clarity of this transition. Other authors, such as Luna et al. \cite{luna2020agile}, provided a practical example in which they operationalized a theory about agile governance by identifying 11 propositions, 24 variables, and 16 hypotheses. This example shows that a well-structured operationalization can produce a large number of testable hypotheses, each focusing on different aspects of the theory. As Stol notes, \emph{``When operationalizing a proposition, it is possible that multiple hypotheses are necessary''} \cite[p. 76]{stol2024teaching}. If the number of propositions is high, the number of hypotheses will also be high. ``The general rule is that a new hypothesis is established each time a different empirical indicator is employed for any one of the units designated in a proposition\footnote{In this work, (empirical indicator = value) and (unit = variable).}'' \cite[p. 215]{dubin:1978}. To avoid this explosion in the number of hypotheses, and following the principle of parsimony, i.e. not multiplying entities unnecessarily\footnote{In words attributed to William of Ockham: ``Pluralitas non est ponenda sine necessitate''}, Dubin \cite{dubin:1978} proposed selecting only the strategic propositions, which are those that describe a state of change in the system, ignoring other propositions that do not provide additional information about the current state.

\section{Proposed Methodological Framework for Theory building}\label{sec:framework}

\begin{figure}
    \centering
\includegraphics[width=0.7\textwidth,keepaspectratio]{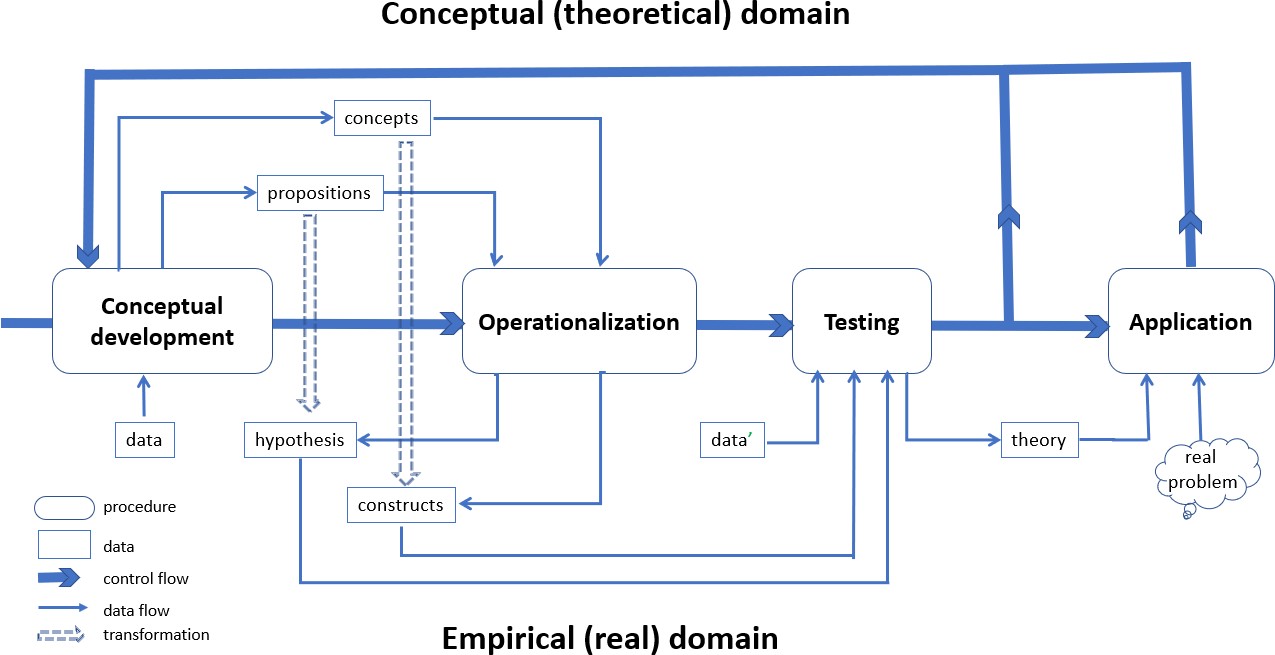}
    \caption{Methodological framework proposed for theory building.}
    \label{fig:method-framework}
\end{figure}

Figure \ref{fig:method-framework} describes the methodological framework proposed in this work for constructing theories. It is an iterative process, similar to Rivard's spiral model \cite{rivard2021theory}, and shares similarities with the phases defined by Lynham \cite{lynham2002}. It also draws on some of the products listed by Hassan, Lowry \& Mathiassen \cite{hassan2022useful}. This framework includes the following phases:

\begin{itemize}
\item
  \textbf{Conceptual development} concerns the conception of relevant concepts and relationships through inductive and abductive processes. In this phase, a qualitative analysis of the data generates a set of concepts and a set of propositions that relate these concepts to each other.
\item
  \textbf{Operationalization} concerns a process by which concepts are converted into constructs by assigning empirical values to their variables, while the propositions generate a set of testable hypotheses.
\item
  \textbf{Testing} refers to the examination of validity through empirical studies to purposefully inform and intentionally confirm or disconfirm/reject the theoretical framework central to the theory. This involves testing the theory, now in the form of constructs and hypotheses, against data from new case studies. It is a deductive process that attempts to determine whether new evidence aligns with the described theory.
\item
  \textbf{Application} concerns the study, inquiry, and understanding of the theory in action, i.e., in real situations in practical disciplines
\end{itemize}

In the first phase (conceptual development), data collection (\emph{data} in Figure \ref{fig:method-framework}) and analysis were conducted using Grounded Theory (GT) procedures in their constructivist version \cite{charmaz:2014}, enriched with collaborative coding and criteria for identifying key concepts \cite{diaz2021GT,gonzalez-prieto_reliability_2021}. This conceptualization phase has already been documented in previous work \cite{diaz2022}. The output products are a set of concepts and a set of propositions that relate those concepts (\emph{concepts} and \emph{propositions} in Figure \ref{fig:method-framework}). Figure \ref{fig:method-framework} shows that the input (\emph{data}) belongs to the empirical (real) domain, while the output products (\emph{concepts} and \emph{propositions}) belong to the conceptual (theoretical) domain.

During the operationalization phase, concepts and propositions are transformed into constructs and hypotheses, respectively. In this case, the transformation is from the theoretical domain to the real domain. In the section \ref{operationalization-1}, we will focus on this process, which is the subject of this paper.

During the testing phase, the aim is to obtain new data (\emph{data'} in Figure \ref{fig:method-framework}) from the real domain and compare them with the hypotheses and constructs obtained so far (see Figure \ref{fig:method-framework}). To carry out this comparison and reduce researcher bias, both the new data and the hypotheses/constructs must be expressed in the same language, and the comparison process must be automatic. The result of this phase is a set of hypotheses that makes up the \emph{theory} (see Figure \ref{fig:method-framework}). During this phase, some hypotheses may be refuted by the analyzed data. This requires a new cycle, beginning with the conceptualization of new evidence; hence the flow from `Testing' to `Conceptual development'. It is therefore an iterative process that gradually refines the theory, bringing its scope of validity closer to the scope of interest. In line with contemporary philosophy of science (e.g., Sober's \cite{sober1999testability} concept of relative testability), we acknowledge that hypotheses cannot be neatly classified as intrinsically testable or untestable in isolation. Within our framework, the testability of a hypothesis is defined relative to the other hypotheses in the network. The testing phase evaluates the entire graph structure against alternative theoretical configurations, assessing how new empirical evidence refutes or supports the interconnectivity of the system as a whole.

The application phase involves using the theory in the real world in order to verify, among other things, its usefulness. In this phase, it is also possible to observe discrepancies between the theory and the problems to which it is applied. This would require new cycles of refinement.

\end{multicols}

\begin{multicols}{2}

It should be clarified that this section has described the methodological framework that guides this research work, the result of which will be a methodology for constructing theories. We are not, therefore, formulating any theory, nor should this work be considered part of a metatheory.

\section{Research Methodology}\label{research-methodology}

The methodology underpinning this research is meta-science \cite{ralph2020empirical} since we address methodological problems associated with theory building in empirical research. Specifically, we address the following research question: How can qualitative theoretical propositions in software engineering be systematically operationalized and optimized to maximize their parsimony and falsifiability? To answer this, throughout the broader research project in which this work is situated, we have employed a Mixed Methods Research (MMR) approach. According to Storey et al. \cite{n12}, MMR \emph{``is a research approach where multiple methods are used to collect, analyze, and integrate both qualitative and quantitative data to address a research problem and produce novel insights''.}

From a pragmatic perspective, the rationale for using MMR in this research is based on the following principles:

\begin{itemize}
\item
  Principle of methodological rationale. In this case, we invoke complementarity. Through constructivist GT, we collect and analyze rich, attribute-laden data, capturing deep semantic insights into the domain where the theory is developed (conceptualization). Through induction, we derive propositions connecting these concepts. Then we transition from the conceptual model to the real model, which includes observable and testable elements. In these subsequent three phases (operationalization, testing, and application), we shift towards an analytical and structural approach with an inference system based on abduction and deduction. Operating within a critical realist paradigm, this formalization allows us to model tendencies and mechanisms objectively, enabling greater consensus and rigor around the developed theory without reducing open empirical phenomena to positivist regularities
\item
  Principle of novel integrated insights. This refers to complementary storytelling. While the qualitative approach generates a rich set of concepts and relationships, the quantitative approach formulates these elements within a formal and rigorous system. In this work, we use logical implication.
\item
  Principle of procedural rigor. Data capture and analysis were conducted following GT procedures \cite{charmaz:2014}, enriched with collaborative coding and criteria for identifying key concept \cite{diaz2022,diaz2021GT,gonzalez-prieto_reliability_2021}. Inductive reasoning was used to derive propositions based on relationships between concepts. The operationalization of concepts into constructs and propositions into hypotheses is explained in the following sections using logical implication to express these hypotheses. To reduce the number of hypotheses, we applied graph theory algorithms that are well- documented and accepted by the scientific community. During testing, we will use algorithms that support automatic deduction (via unsatisfiability detectors) to validate or refute the hypotheses forming the theory by contrasting them with hypotheses derived from new empirical data.
\end{itemize}

According to Storey et al. \cite{n12}, the design of this research is Exploratory Sequential (planned), with transformation- based integration. The qualitative (quaL) approach is applied in the first of the four phases, and the quantitative (quaN) approach in the remaining three. The data obtained through quaL is transformed into quaN data. The inference scheme follows induction and abduction in the first two phases and deduction in the last two.

\subsection{Epistemological Positioning: Critical Realism}
To coherently articulate the four phases of the proposed methodology, this work is conceptually positioned within Roy Bhaskar's Critical Realism \cite{bhaskar1979philosophy,bhaskar2008realist}. This philosophical stance provides a framework capable of integrating interpretive and analytical elements without reducing reality either to social constructions or to observable empirical regularities.

As established in our framework, during the conceptualization phase (phase 1), which is interpretive and focused on the actors' constructions of meaning, initial concepts and propositions emerge directly from the discourses and social constructions of the studied actors using Grounded Theory (GT) procedures in their constructivist version \cite{charmaz:2014}. In addition, inductive reasoning was applied to derive propositions as relationships between concepts.

Subsequently, during the operationalization phase (phase 2), these concepts and propositions are transformed into theoretical constructs and formal hypotheses. Characterizing this phase from a critical realist perspective, we note that since software engineering phenomena function as open empirical systems (as detailed in Appendix \ref{app:a1-main}), operationalization here does not aim to establish metric or covariance-based scales of continuous latent variables typical of psychometrics or econometrics. Instead, it refers to a syntactic and logical specification of hypotheses, converting qualitative propositions into structured statements of logical implication to systematically map real-world mechanisms.

Finally, as the theory enters the testing (phase 3) and application (phase 4) stages, adopt an analytical formalism. In these stages, graph theory and formal logic are employed not as positivist measurement tools, but as a structural scaffolding to map the logical relationships among the operationalized hypotheses and to allow for the evaluation and eventual Popperian refutation of the formulated hypotheses. In the testing phase, the goal is to contrast the theory, expressed as a set of hypotheses H, with new empirical evidence H' to verify deductive alignment or falsification. Finally, the application phase assesses the suitability of the theory within its empirical domain, where fuzzy logic may be utilized to capture the nuances of practical reality. In this manner, the mathematical formalism is placed at the service of a deep and rigorous understanding of a stratified reality.

\section{Operationalization}\label{operationalization-1}

This section presents our formal methodological proposal for how the operationalization phase should be systematically executed in software engineering theory building. The operationalization phase is described from three dimensions. First, the procedures for operationalizing are described. Second, the same procedures are formally specified. Finally, the usefulness of the proposed procedures will be demonstrated with a simple case study.

The need to formalize this process is due to the strategy that will be followed during the testing phase. To avoid bias in interpreting whether the new evidence provided during testing aligns with the hypotheses resulting from the operationalization phase, we have decided to automate this comparison. To automate it, a minimum level of formalization is necessary for expressing both the hypotheses generated during operationalization and the new hypotheses derived from testing.

In a constructivist stance, bias is inherent. However, it is important to recall that the ontological/epistemological approach in this phase is critical realism, where minimizing researcher bias is essential to safeguard the structural reliability of the formalized generative mechanisms. The formal definitions introduced here are required only for the automated reasoning process; researchers using the methodology do not need to manipulate these formalisms directly.

\emph{Example}. This case study uses the results generated during the conceptualization phase carried out to build a theory on ``the structure of IT departments and teams when software-intensive organizations adopt DevOps'', which is described in Díaz et al. \cite{diaz2022}. The complete study includes 8 concepts (see appendix \ref{app:d1-concepts}) and 28 propositions (see appendix \ref{app:d2-propositions}) obtained during the conceptualization phase, and 5 constructs, 24 variables, 12 universes, and 28 hypotheses generated during the operationalization phase (see appendixes \ref{app:d3-constructs}, \ref{app:d4-universes} and \ref{app:d5-hypotheses}). These constructs, variables, and hypotheses were handled by the automated tool developed in this work and described in Appendix \ref{synt-a-software-tool-for-automating-the-theory-synthesis-process}, thereby ensuring the scalability of the approach. The tool reduced the final number of hypotheses by 31\%.

The operationalization of a theory addresses two activities: (i) operationalizing theoretical concepts into constructs; and (ii) operationalizing theoretical propositions into empirically testable hypotheses. After these two activities, we add a third activity of synthesis, which aims to reduce the number of generated hypotheses to maintain the principle of parsimony.

\subsection{Operationalizing theoretical concepts into constructs}\label{operationalizing-theoretical-concepts-into-constructs}

A construct represents a phenomenon that is not directly measurable. Instead, it must be inferred from other elements that can be directly observed. These elements are called variables and are quantified using some kind of measuring instrument. We now describe some aspects that, from our point of view, should be taken into consideration.

\begin{itemize}
\item
  A construct can encompass several concepts. It is necessary to analyze the set of concepts derived from the conceptualization phase and derive constructs from them
\item
  A clear and precise definition of the construct is fundamental for its reliability. This definition must be operational in the sense that it must indicate the terms in which it can be measured. The constructs can be unidimensional if they have only one variable and one dimension associated with them (in the style of single-operation concepts) or multidimensional if they have several variables associated with them. In the case of unidimensional constructs, it is necessary to analyze whether they can generate an under representation of the concept.

  \item
    A construct must have associated scope conditions; that is, the context in which this construct operates \cite{suddaby2010construct}.
  \item
    Each of the variables (dimension) associated with a construct must be inferred from the analyzed data or derived from abductive reasoning. Each variable has an associated type that represents the universe in which it takes values.
  \item
    Additionally, the relations between values of the variables must be determined in order to allow comparison. These relations must be decided for each of the types of variables determined in the previous step.
  \end{itemize}

The selection of a propositional logic system with variables, fully developed in Appendix \ref{propositional-logic-with-variables}, provides the appropriate mathematical framework to formalize these variables and types. While higher-order frameworks (such as predicate logic) offer greater semantic granularity, they introduce computational complexity and potential undecidability for automated validation frameworks. Propositional logic provides the exact level of formal precision required to model the structural dependencies between constructs as clear directed edges. This constraint is critical because it directly enables our automated graph-theoretic algorithms to systematically eliminate deductive redundancies without losing semantic alignment in open systems. Generally speaking, as the outcome of this step in the operationalization procedure of concepts, we must obtain a language $\mathcal{L} = \{\mathcal{U}, \mathcal{V}, \mathcal{F}, \mathcal{R}\}$, where:

\begin{itemize}
\item
  $\mathcal{U} = (U_{1}, \ldots, U_{m})$ are the
  \emph{types} (referred to as \emph{universes} in this context), and $m$ is the total number of types considered in the theory.
\item
  $\mathcal{V} = (V_{1}, \ldots, V_{n})$ are the
  \emph{variables}, where $V_{i}$ denotes the collection of
  variables with type in the type $U_{i}$.
\item
  $\mathcal{F} = (F_{1}, \ldots, F_{f})$ are the
  \emph{functions} considered over values of types, where $F_i$ denotes the set of functions with values in the type $U_i$.
\item
  $\mathcal{R} = (R_{1}, \ldots, R_{r})$ are the
  \emph{relations}, with $R_{i}$ the relations on the
  type $U_{i}$.
\end{itemize}

With these considerations in mind, a table is constructed for each construct, as shown in Appendix \ref{app:d3-constructs}. In addition, another table is generated, as shown in Appendix \ref{app:d4-universes}, which describes the different types/universes characterized by their values, the variables belonging to that universe, functions and relationships.

For example, Table \ref{tab:d3-construct-interaction} (Appendix \ref{app:d3-constructs}) describes the `Interaction' construct, which is derived from the concepts of `Collaboration' and `Communication' (the authors have chosen to combine these two concepts into a single construct), thereby making these concepts dimensions of the construct (which is, therefore, multidimensional). The operational definition of the construct and the context to which it applies (scope conditions) are also specified. For each dimension, it is indicated whether this variable has been derived from data analysis or obtained through abductive reasoning. Furthermore, the type (single-valued -- scalar -- or multi-valued -- collection --) is specified, as well as the values that each variable can take.

Meanwhile, Table \ref{tab:universes} describes the types (universes) for this example. For each type, the possible values are specified, along with the variables belonging to this domain (indicating their name, acronym and the construct with which they are associated) and the functions and relations associated with this type.

Therefore, the associated language $\mathcal{L} = \{\mathcal{U}, \mathcal{V}, \mathcal{F}, \mathcal{R}\},$ is specified as
\begin{align*}
\mathcal{U} & = (Tty, Tau, Tag, Tre, Tbo, Tma, Tcm, Tpr,\\
&\qquad Tva, Tbp, Tat, Tco),
\end{align*}
where:
\begin{align*}
Tty &= \{\text{product-team}, \text{bridge-team}, \text{enabler-team},\\
&\qquad \text{development-team}, \text{operations-team}\},\\
Tau &= \{\text{dependent}, \text{self-organization}\},\\
Tag &= \{\text{local-optimization}, \text{product-thinking}\},\\
Tre &= \{\text{Interval }[0,10]\},\\
Tbo &= \{\mathsf{true}, \mathsf{false}\},\\
Tma &= \{\text{product-management}, \text{project-management}\},\\
Tcm &= \{\text{small-frequent}, \text{large-rare}\},\\
Tpr &= \{\text{collection }(\text{customer-centric-action},\\
&\qquad \text{end-to-end-responsibility},\\
& \qquad \text{automate-everything})\},\\
Tva &= \{\text{collection }(\text{communication}, \text{collaboration}, \\
&\qquad\text{transparency})\},\\
Tbp &= \{\text{collection }(\text{Continuous-}\allowbreak\text{integration},\\
& \qquad
\text{continuous-testing},
\text{continuous-delivery},\\ & \qquad\text{continuous-deployment},
\text{continuous-monitoring},\\& \qquad \text{continuous-improvement},\\
&\qquad
\text{infrastructure-}\allowbreak\text{as-}\allowbreak\text{code})\},\\
Tat &= \{\text{automated-}\text{infrastructure-}\allowbreak\text{management-interface},\\
&\qquad\allowbreak
\text{automated-}\allowbreak\text{application-}\allowbreak\text{life-cycle-}\\
&\qquad\allowbreak\text{management-interface}\},\\
Tco &= \{\text{(Daily, High)},\allowbreak \text{(Eventual, Low)},\\
&\qquad \text{(Daily, Low)},\allowbreak \text{(Eventual, High)}\}.
\end{align*}

For the variables, functions and relations, we have
\begin{align*}
  \mathcal{V} &= (\{Tt\}, \{Au\}, \{Ag\}, \{Rs, Ks, Ss\},\\
&\qquad\{Bm, Rd, Cf, So, Mv, Tw, Sc, Aa, Ai, Ps, Cm\},\\
&\qquad \{Me\}, \{Ch\}, \{Pr\}, \{Va\}, \{Bp\}, \{Pl\}, \{Co\}),\\
\mathcal{F} &= (\varnothing, \varnothing, \varnothing, \varnothing, \varnothing, \varnothing,\varnothing, \varnothing, \varnothing, \varnothing, \varnothing, \varnothing),\\
\mathcal{R} &= (\varnothing, \{>\}, \{>\}, \{>\}, \{=\}, \varnothing, \{>\}, \varnothing, \varnothing, \varnothing, \varnothing, \{=, >\}).
\end{align*}

A particularly important issue during this operationalization phase concerns quantification. It is not the purpose of this paper to develop a general theory of measurement. The discussion provided here is limited to the aspects of quantification that are necessary for the operationalization process, while acknowledging that the detailed technical treatment of measurement lies outside the scope of this study. The selection of variables to represent a concept and the assignment of possible values to those variables (to generate a construct) will condition the resulting hypotheses. The problem of quantification is multifaceted and specific to each domain, and researchers will have to address it during the development of their theory. Consequently, quantification cannot be approached as a general problem or resolved by means of a uniform procedure applicable to any theory. Consistent with the views expressed in the measurement and theory-building literature \cite{flake2020measurement,furr2021psychometrics,goertz2020social,suddaby2010construct}, the values that a variable can take necessarily depend on how the construct has been defined, the empirical context in which it is applied, and the type of statements that the theory aims to formulate. For this reason, no methodology can anticipate all possible domains in advance. What can be offered instead is a framework that allows for explicit justification of the decisions made during quantification. One recommendation is that the researcher justify the reasons that led them to declare the selected variables and their associated values (including threshold values) or the reasons for combining concepts into a single entity. Variables and values can be justified by data obtained empirically through interviews, surveys, ethnography, focus groups, and other relevant sources. This increases traceability and reliability. Issues related to the validity of the operationalization process, including potential sources of bias and limitations in variable/value selection, are discussed in Section \ref{threats-to-validity-reliability-and-limitations}.

\subsection{Operationalizing theoretical propositions into empirically testable hypotheses}\label{operationalizing-theoretical-propositions-into-empirically-testable-hypotheses}

A primary aspect to consider is the way in which hypotheses are expressed. Since early research on theory formulation originated in the field of social sciences (sociology, psychology, anthropology, etc.), causal reasoning is commonly associated as the standard approach for expressing hypotheses. However, despite its widespread use, this research does not describe hypotheses in causal form (see Appendix \ref{app:a1-main}). This is one of the key differences between our work and the approaches described in Kanaparan \& Strode \cite{Kanaparan:2021} and Luna, Marinho \& de Moura \cite{luna2020agile}. Both Kanaparan \& Strode \cite{Kanaparan:2021} and Luna et al. \cite{luna2020agile} operationalize their theories based on the existence of dependent, independent, and mediating variables, within a causal reasoning framework. Current approaches to operationalization are based on causal relationships. It should be noted that causal claims, even in their weakest statistical or counterfactual forms, entail ontological commitments about the dependencies that connect cause and effect. In contrast, the logical implications used in this work do not assert any mechanism or metaphysical structure; they are purely formal constructs introduced for operational and analytical purposes. From this perspective, hypotheses consist of two parts: an antecedent and a consequent, where the latter is the logical outcome of the former. That is, the antecedent is not the cause, nor is the consequent the effect (as it would be in a causal relationship).

A second aspect to consider is the possible explosion of the number of hypotheses that will be generated. We recall that a new hypothesis is established each time a different empirical indicator is employed for any of the units (variables) designated in a proposition. In order to meet the parsimony criterion, one may choose to eliminate non-strategic propositions. Note that this is a subjective judgment and will introduce a bias in the resulting theory. Another strategy is to obtain the canonical set of hypotheses, which is the minimum and necessary set that fully describes the system. In Section \ref{synthesis-from-the-theory}, this last strategy and a procedure to automate it are formally defined.

Prior to this, it is essential to outline the procedure for formalizing
hypotheses so that they can be expressed as logical implications. The
formalization will again be driven by the language of propositional
logic with variables, as described in Appendix~\ref{propositional-logic-with-variables}. Roughly speaking, in
this logic, the formulas are created in two levels. In the first level,
we create the literals by considering strings of the form $P(t_{1}, \ldots, t_{n})$, where
$(t_{1}, \ldots, t_{n})$ are terms created by combining variables, constants and functions, and $P$ is a relation of arity $n$.
In the second level, we create formulas by combining literals and simpler formulas using the logical connectors $\vee$ (``or''), $\wedge$
(``and''), $\to$ (``implies''), $\leftrightarrow$ (``if and only if'') and $\neg$ (``not'').
In this manner, the hypothesis formalized in this step of the operationalization process will lead to a collection of formulas $T = \{\varphi_{1}, \ldots, \varphi_{n}\}$, usually referred to as
the theory.

For example, propositions P1, P2, P6 and P10 described in Appendix \ref{app:d2-propositions} are converted into formulas of the language $\mathcal{L}$. In particular, we model these propositions as follows:

\begin{itemize}
\item
  \emph{P1: ``A team culture based on responsibility/ownership sharing enables collaboration''.} If we consider that a company has achieved a culture based on ownership sharing when $\text{Rs} > 5$, and that collaboration exists when we have more than (\emph{Eventual, Low}) collaboration, then the resulting formula is
$$
    \varphi_1 = [\text{Rs} > 5] \longrightarrow [\text{Co} > (\text{Eventual},\text{Low})].
$$

Note that the above considerations must come from the operationalization process and be justified; they cannot be arbitrary. In this case, these values ``$> 5$'' and ``$> (Eventual, Low)$'' have been derived from the interviews analyzed (using Grounded Theory) during the conceptualization phase.

\item
  \emph{P2: ``Promoting collaboration reduces organizational silos/conflicts''.} With the same interpretation that collaboration exists when it holds $Co > (\text{Eventual}, \text{Low})$, this formula translates into ``If there is collaboration, then there are no silos'', that is
$$
\varphi_2
= [\mathrm{Co}>(\text{Eventual},\text{Low})]
\longrightarrow
\neg[\mathrm{So}=\mathrm{true}].
$$

\item
  \emph{P6: ``[...] if there are already separate roles, responsibilities are very clear and collaboration is not fostered or promoted''.} Observe that the consequence of the implication is negated (no collaboration is expected). Therefore, the formula obtained is the following
$$
\varphi_3
= [\mathrm{Rd}=\mathrm{true}]
\longrightarrow
\neg[\mathrm{Co}>(\text{Eventual},\text{Low})].
$$
\item
  \emph{P10: ``Responsibility/ownership sharing reduces organizational silos/conflicts''.} Again, understanding that ownership sharing occurs when $\textrm{Rs} > 5$, we model this proposition as
$$
\varphi_4
= [\mathrm{Rs}>5]
\longrightarrow
\neg[\mathrm{So}=\mathrm{true}].
$$
\end{itemize}

With these propositions, the theory representing the hypotheses of the case study is:
$$
    T = \{\varphi_1,\varphi_2,\varphi_3,\varphi_4\}
$$

\subsection{Synthesis from the theory}\label{synthesis-from-the-theory}

As the outcome of the process of Sections \ref{operationalizing-theoretical-concepts-into-constructs} and \ref{operationalizing-theoretical-propositions-into-empirically-testable-hypotheses}, the operationalization procedure has determined a language $\mathcal{L}$ and a finite theory $T$ of formulas in the language $\mathcal{L}$, representing the hypotheses formulated by formalizing the propositions in the theory.

However, in general, this theory $T$ will have a limited scope due to two somehow opposite phenomena:

\begin{enumerate}
\def\labelenumi{(\arabic{enumi})}
\item
  $T$ may not contain all the hypotheses that can be logically deduced from it.
\item
  $T$ may contain redundant hypotheses that can be removed without affecting the capability of the theory.
\end{enumerate}

Fortunately, both problems can be addressed using computational techniques, as described in \ref{automatic-deduction}. Explicitly, to address problem (1), using the Davis-Putnam algorithm (see Algorithm \ref{alg:davis-putnam}), we can easily check whether a formula $\varphi$ follows from theory $T$ or not. In the same way, as described in Algorithm \ref{alg:minimal-theory}, it is possible to extract a minimal subtheory $T_0$ of $T$ with the same logical capacity and the fewest necessary hypotheses.

Furthermore, in the case where the formulas of the theory $T$ have a simple form, we can even use graph-theoretic algorithms to address problems (1) and (2), as described in \ref{automatic-deduction-in-implication-theories}. In particular, we need $T$ to be an implication theory, which means that all the formulas in $T$ are of the form
$$
    \ell \to \ell',
$$
where both $\ell$ and $\ell'$ are literals or negations of literals. These theories, despite being very specific, tend to appear in qualitative research, since each of the propositions formalized concern typically two constructs related by an implication (``A'' implies ``B'').

In this case, the theory $T$ can be described as a directed graph $\mathcal{G}_T$, as described in Appendix \ref{automatic-deduction-in-implication-theories}, whose nodes are the possible literals or negation of literals appearing in the theory and whose edges are the implications presented in the theory and their counter-reciprocals.

It is crucial to note that the graph-theoretic algorithms described in Appendix \ref{automatic-deduction-in-implication-theories} operate strictly at the level of syntactic and logical parsimony. Unlike statistical models such as Structural Equation Modeling (SEM) or Bayesian Causal Directed Acyclic Graphs (DAGs), where removing a direct path because an indirect path exists would violate the coexistence of direct and indirect effects and alter model fit, our transitive reduction functions purely as an optimization tool for theoretical statements. It removes deductive redundancy within a network of logical implications to achieve a more parsimonious expression of the theory, without making any metaphysical claims about the elimination of multi-path causal dependencies in the empirical world.

It is not strictly necessary to understand these graph-based formalisms. In this work, a software tool (see Appendix \ref{synt-a-software-tool-for-automating-the-theory-synthesis-process}) has been implemented that is capable of performing all these transformations automatically. Given the language L defined at the end of Appendix \ref{app:d4-universes} and the hypotheses defined in Appendix \ref{app:d5-hypotheses}, the tool, by applying graph algorithms, generates the minimal set of hypotheses that represents the theory T (Appendix \ref{synt-a-software-tool-for-automating-the-theory-synthesis-process} describes this set as well as the redundant hypotheses that have been removed).

\section{Results}\label{results}

This paper has described the definition and implementation of a crucial phase in the creation of substantive theories in empirical research, that is, operationalization. The operationalization phase has been formally described from three points of view: specification, implementation, and application. The novelty is relevant not only in the formal specification but also in the design of the procedures and their interrelationships.

During the operationalization of concepts, we have developed criteria to generate the constructs and to describe the variables that characterize them and the empirical values that these variables can take. During the operationalization of propositions, two novel dimensions of this work should be highlighted. First, we propose and explain the use of logical implication instead of causality. Secondly, we design a set of strategies to maintain the parsimony of the theory. In this regard, the proposal to generate the canonical set of hypotheses has not been previously applied in the literature. Reducing the number of hypotheses has significant strategic advantages. These include: i) optimization of resources: it optimizes empirical test planning by identifying the core, independent relationships of the theory; and ii) maximum conceptual clarity: it forces the team to identify the core, non-redundant relationships in their theory. A software tool called \textbf{SynT} has been developed which, given the language of a theory, allows its constructs and hypotheses to be entered and calculates the minimal theory by eliminating redundant hypotheses.

The quantitative results presented in the case study serve as a formal evaluation of the analytical utility and feasibility of our proposal. In methodological and meta-scientific research, demonstrating that a newly developed formal artifact can successfully process a complex, real-world network, reducing an unmanageable set of qualitative propositions to a minimized, optimized core without losing semantic validity, constitutes a rigorous and standard form of validation in open systems. Specifically, these metrics illustrate two critical aspects of our methodology. First, they demonstrate the elimination of semantic and deductive redundancy; the mathematical reduction strips away the extraneous pathways that human reasoning accumulates when dealing with complex qualitative narratives. Second, and most importantly, this example illustrates a drastic reduction in cognitive load and field research cost. Instead of forcing the researcher to blindly design empirical tests for an unmanageable explosion of isolated verbal hypotheses, the minimized core derived from the graph transformations provides a parsimonious, visualizable roadmap of the absolute essential dependencies that require validation in the open system.

\section{Threats to Validity \& Reliability and Limitations}\label{threats-to-validity-reliability-and-limitations}

Criteria for judging the quality of research designs are essential to establish the validity (i.e., the accuracy of the findings), and the reliability (i.e., the consistency of procedures and the researcher's approach) of most empirical research  \cite{creswell:2003,yin:2017}. These criteria must address two domains: the methodological and the applicative, both referring to the operationalization phase, which is the focus of this work.

\subsection{Threats to Methodological Design}\label{threats-to-methodological-design}

The methodological proposal presented here combines several phases with distinct epistemological and logical foundations: an inductive conceptualization based on constructivist Grounded Theory \cite{charmaz:2014}, an abductive operationalization under a critical realist stance, and testing and application phases guided by deductive principles aimed at structural falsification in open systems. While this integration offers a robust perspective for theory building and evaluation in the domain of software engineering, it also introduces significant challenges that must be explicitly considered. Some of these include:

\begin{itemize}
\item
  Epistemological coherence. A primary threat arises from the combination of epistemological paradigms traditionally considered disparate. The use of a constructivist logic for the initial generation of concepts and propositions may appear to be in tension with their subsequent quantification and empirical verification. However, this transition is justified within a Mixed Methods Research (MMR) approach (see Section \ref{research-methodology}) with transformation-based integration, where the qualitative (quaL) approach is applied in the first of the four phases, and the quantitative (quaN) approach in the remaining three. The data obtained through quaL is transformed into quaN data.

  \item
    Mixed inference. All three forms of inference have been used. Section \ref{research-methodology} justifies the use of induction in the conceptualization phase, abduction in the operationalization phase (to select plausible variables), and deduction in the testing and application phases. The coexistence of these distinct inference mechanisms may generate ambiguities regarding how theoretical results are articulated with empirical data. In particular, the transition from inductively emergent propositions to abductive operationalization requires justification for how observable variables are selected to represent abstract concepts. As indicated at the end of Section \ref{operationalizing-theoretical-concepts-into-constructs}, this is a quantification problem that falls outside the scope of this work.
\end{itemize}

Validity criteria in research using MMR depend on the type of design adopted by the researcher. As indicated in Section \ref{research-methodology}, this study is based on an Exploratory Sequential (planned) design with transformation-based integration. According to Creswell \cite{n13}, this type of design presents potential validity threats, as shown in Table \ref{tab:validity-threats-esd}, along with strategies to mitigate them.

\begin{center}
    \centering
    \captionof{table}{Validity threats for Exploratory Sequential Design (adapted from Creswell \cite{n13}).}\label{tab:validity-threats-esd}
\begin{tabularx}{\columnwidth}{|m{0.35\columnwidth}|m{0.55\columnwidth}|}
\hline \textbf{Validity Threats}
& 
\textbf{Strategies to Minimize Threats} \\\hline\hline
(T1) Not building the quantitative feature based on the qualitative
results & Make explicit how each qualitative finding is used to inform
the development of specific elements of the quantitative feature \\\hline
(T2) Not developing rigorous quantitative features & Use systematic
procedures to design the quantitative feature (e.g., apply sound
psychometric instrument design steps or pilot test intervention
materials) \\\hline
(T3) Selecting participants for the quantitative test that are the same
individuals as the qualitative sample & Use a large sample of
individuals for the quantitative phase who are different from those in
the qualitative sample \\\hline
    \end{tabularx}
\end{center}

Hence, regarding threat (T1), Section \ref{operationalizing-theoretical-concepts-into-constructs} provided guidelines on how to generate constructs from concepts. As for the transition from propositions to hypotheses, Section \ref{operationalizing-theoretical-propositions-into-empirically-testable-hypotheses} presented the proposed formalization for this process. Since the synthesis process (Section \ref{synthesis-from-the-theory}) is a formal optimization of the product obtained within the operationalization phase, it does not introduce new empirical data threats. However, a potential conceptual threat is the misinterpretation of our transitive reduction as a tool for statistical causal modeling. To mitigate this threat, we explicitly delimit the scope of our graph transformations: they are designed for propositional logic systems with variables (Appendix \ref{propositional-logic-with-variables}) to minimize statement explosion. They do not compete with covariance-based methods or causal graphs (e.g., \cite{furia2023towards}), which model data-driven effects. If a researcher intends to subsequently translate the resulting minimized theory into an empirical structural equation model, the links removed by our syntactic reduction must be reintroduced to test for concurrent direct and indirect effects in the open system. With respect to threat (T2), the entire process is formalized using propositional logic with variables (see Appendix \ref{propositional-logic-with-variables}). Furthermore, the proposed protocol for obtaining the canonical set of hypotheses is based on graph theory, a formalism widely accepted by the scientific community. Finally, threat (T3) may be relevant during the testing phase, but not in this operationalization phase. In any case, this threat can be eliminated by ensuring that the data used to repeat the conceptualization process are different from those used to generate the original hypotheses (see Section \ref{research-methodology}).

\subsection{Threats to Application}\label{threats-to-application}

These threats must be managed by the researcher applying this methodology in order to mitigate their impact on the outcomes of the investigation. Although these threats do not pertain to the methodology itself, as those described in Section \ref{threats-to-methodological-design}, they are presented here as supplementary information that researchers should consider when applying this methodology. Given that the epistemological stance adopted for the operationalization phase is critical realism, focusing on the structural dependencies of open systems, the validity threats are outlined below.

\emph{Validity. Construct Validity.} Several threats may introduce bias in the process of constructing both constructs and hypotheses. Regarding constructs, different researchers may derive different sets of constructs from the same set of concepts, characterizing them with different variables that may be assigned different values. This bias affects reliability, as different researchers may generate different hypotheses from the same input data and procedures. Additionally, the procedures described to reduce hypothesis explosion, such as eliminating non-strategic propositions, may result in different sets of hypotheses depending on the researcher. Again, these biases impact reliability. To mitigate these threats, multiple researchers can independently perform the same task and then apply an inter-rater agreement protocol.

\emph{Internal Validity}. The main threat lies in the abductive process through which new (plausible) variables may be introduced to complete the characterization of a construct, especially in the case of unidimensional constructs. Different researchers may add different variables, introducing bias that affects reliability.

\emph{External Validity}. The outcome of the operationalization process is a set of hypotheses valid within the domain in which they were generated. That is, they form part of a substantive theory and are therefore not generalizable.

\emph{Reliability.} Reliability is largely dependent on the biases introduced by researchers. As noted in the previous section, reliability may be compromised by the various biases incurred during the practical implementation of the operationalization process.

\subsection{Limitations}\label{limitations}

The initial conceptualization phase, being based on constructivist Grounded Theory, may produce highly contextual propositions whose scope could limit the external validity of subsequent phases. This is not a risk per se, but rather a limitation of the proposed methodology for theory construction.

\section{Conclusion \& future works}\label{conclusion-future-works}

This paper presents a set of procedures to systematize the operationalization phase in theory building. The contribution presented in this paper is part of a larger research initiative focused on theory building \cite{diaz2021GT,gonzalez-prieto_reliability_2021} and its application to advance knowledge in the fields of DevOps \cite{diaz2022,lopez2021}, and edge computing \cite{perez2022}. Specifically, this paper illustrates the operationalization phase using the theory on DevOps team structure that was previously described in \cite{diaz2022}.

This research outlines a comprehensive process for operationalization, proposing a set of procedures to operationalize both concepts and propositions. Innovations presented include methods for expressing hypotheses (logical implications) and strategies to mitigate the proliferation of hypotheses through the use of a canonical set of hypotheses. By separating the semantic domain of qualitative interpretation from the syntactic domain of formal variables, this approach, for the first time, allows a qualitative theory to be converted into a computable model that can be automatically analyzed and optimized. Operating under a critical realist paradigm, this formalization does not attempt a superficial positivist quantification, but rather functions as a vital cognitive support tool to manage the complexity of generative mechanisms in open systems. By mathematically resolving deductive redundancies through graph transformations, the methodology preserves the principle of parsimony and prevents the cognitive overload caused by hypothesis explosion. Ultimately, this structural scaffolding provides researchers with a minimized, rigorous roadmap of core dependencies, which will save weeks or months of unguided empirical testing work in the field.

Concurrently, we are working on strategies to address the testing of the theory. This work has chosen to describe the theory in the language of propositional logic with variables, anticipating that this will facilitate the testing phase and allowing its instrumentalization through logic programming. This phase is a purely deductive process in which new evidence must be validated or refuted against the existing theory. To achieve this, it is essential that the evidence is expressed in the same language as the theory itself. But the propositional logic assigns only two truth values to hypotheses: true or false. This circumstance does not usually occur in the empirical world, where there is a variety of ranges and nuances. Therefore, the aim is to carry out several rounds of testing in the DevOps domain (with different empirical data) to determine the degree of certainty of each of the hypotheses that form the theory developed in this work. This degree of certainty can be modelled with fuzzy logic, which is of vital importance for the application phase.

Future studies should explore this methodology in a variety of empirical settings to assess its robustness and general utility. In fact, a study has already been launched to develop a theory about the impact of AI on teaching/learning and assessment processes in higher education.

\subsection*{Acknowledgments}\label{acknowledgments}
No funding was received.

\subsection*{Disclosure of interest}\label{disclosure-of-interest}
There are no relevant financial or non-financial competing interests to report.

\subsection*{Data Availability Statement}\label{data-availability-statement}
All data generated or analyzed during this study are included in this article.

\bibliographystyle{abbrv}   
\bibliography{bibliography}

\clearpage

\end{multicols}

\appendix

\section{Theoretical framework of theory construction}\label{theoretical-framework-of-theory-construction}

\subsection{Causality and the pragmatic role of logical implication}\label{app:a1-main}

The purpose of this appendix is not to dismiss causal reasoning, which remains a central and productive tool for theory building across the social and behavioral sciences. Instead, the aim is twofold: i) to document recurrent philosophical and methodological limitations associated with causal explanations in open empirical systems; and ii) to justify the pragmatic use of logical implication (rather than causality) as the preferred representational form in the context of theory testing in empirical sciences. It should also be noted that causal claims, even in their weakest statistical or counterfactual forms, entail ontological commitments about the dependencies that connect cause and effect. In contrast, the logical implications used in this work do not assert any mechanism or metaphysical structure; they are purely formal constructs introduced for operational and analytical purposes.

Since early research on theory formulation originated in the field of social sciences (sociology, psychology, anthropology, etc.), causal reasoning is commonly associated as the standard approach for expressing hypotheses. According to Jaccard and Jacoby \cite[p. 151]{jaccard2020}, \emph{``Causal thinking, and the causal modelling that often goes with it, is probably the most prominent approach to theory construction in the social sciences''}. Causal analysis identifies relationships between variables with the premise that a change in one will cause a variation in the other.

The validity of causality has been questioned by many philosophers of science. Hume \cite{hume1975} argued that it is impossible to ever demonstrate that changes in one variable produce changes in another. Russell \cite{russel1931} argued that causality can be established unambiguously only in a completely isolated system (such as mathematics), and empirical science is not. In other words, when faced with empirical evidence, we may infer a cause-effect relationship: event $A$ causes effect $B$, but unless the system is closed, we cannot categorically assert that effect $B$ derives solely and exclusively from event $A$. Empirical software development environments are open systems in which human, organizational, and technological factors interact in complex and often unpredictable ways. Variables such as developer expertise, team coordination, tool usage, task complexity, stakeholder pressure, and organizational processes cannot be fully isolated, controlled, or stabilized across contexts. As a result, empirical software engineering phenomena exhibit variability and context-dependence rather than deterministic regularity, making strict causal determinism unattainable in this domain. Instead, theory building relies on (a) probabilistic causality (e.g., variable $W$ increases the likelihood of outcome $X$), (b) counterfactual reasoning (e.g., interventions on factor $W$ would lead to different expected outcomes), and (c) mechanistic or generative causality only when specific process mechanisms can be articulated (e.g., mediating variables such as coordination or expertise). These forms of causality allow theory builders to express explanatory relationships without assuming closed-system determinism.

Dubin \cite{dubin:1978} argued that many relationships that are labelled causal are not; they are sequential interactions: phenomenon $A$ is observed to occur temporally after another $B$; and this does not imply that $A$ causes $B$. For example, an alarm clock that goes off every morning just before sunrise cannot be said to be the cause of the sun rising, even though the two events are closely linked \cite{jaccard2020}.

Finally, Popper \cite[p. 39]{n10} states: \emph{``The `principle of causality' is the assertion that any event whatsoever can be causally explained---that it can be deductively predicted.''} With this definition, Popper emphasizes that the principle is not falsifiable and, therefore, neither adopts nor rejects the principle of causality. Instead, he excludes it from the realm of science and places it within the domain of metaphysics.

For the purpose of testing theories, what matters is that the theory generates testable implications. If $A$ implies $B$ ($A \to B$), then observing $\neg B$ and $A$, falsifies the implication. This is the core of Popperian testability and the foundation of scientific logic. Within the framework of critical realism, these logical implications act as formal representations of generative mechanisms and tendencies operating in open systems. Unlike deterministic causal claims, these logical implications do not require closed systems or assume that X linearly produces Y in every context. Instead, they map the structural and conditional dependencies of the hypothesized mechanisms. This makes logical implication an opportunistic but powerful tool in empirical environments, such as software engineering, where causal isolation is impossible, allowing researchers to formalize and test the structural validity of a theory without falling into deterministic reductionism

\subsection{Theory}\label{app:a2-main}

\subsubsection{What is a theory?}\label{app:a1-theory} There are multiple definitions of theories, each of which elaborates on some conceptual or structural aspect of the theory. Some of these are described below. Weick \cite{weick1989theory} considers a good theory to be a plausible theory, needing to be interesting, novel, a source of unexpected connections, high in narrative rationality, aesthetically pleasing, and corresponding to presumed realities. This highlights the importance of uncovering connections and relationships not directly observable in the empirical world.

From a management and organizational perspective, Suddaby \cite[p. 407]{n7} defines a theory as \emph{``simply a way of imposing conceptual order on the empirical complexity of the phenomenal world.''}

Similar to Suddaby, insofar as theory must explain real-world phenomena, Sandberg and Alvesson argue that despite the vagueness of definitions, there is more convergence than divergence: \emph{``they converge on the idea of `theory' as conceptual knowledge that aims to explain phenomena''} \cite[p. 488]{n8}.

In the context of software engineering, Johnson and Ekstedt \cite[p. 182]{n11} define a theory as: \emph{``a theory of software engineering is thus a collection of sentences, propositions, statements or beliefs, etc., about software engineering, and their logical consequences. We are aware that some definitions pose additional requirements on theories, such as being internally consistent, well-substantiated by observation, clearly delimited, formalized, falsifiable, sufficiently general, connected, or able to make sufficiently accurate predictions.''}

Gregor \cite{gregor:2006} identifies multiple perspectives on theory, including (i) theory as statements prescribing how something should be done in practice; (ii) theory as a lens for viewing or explaining the world, and (iii) theory as testable statements of relationships among constructs. She emphasizes that a minimal definition of theory must meet three criteria: (1) constructs must be identified; (2) relationships among these constructs must be specified; and (3) these relationships must be falsifiable (that is, able to be tested).

Finally, Sch\"oning \cite[p. 67]{n9} defines a theory from a logical perspective as: \emph{``A theory is a non-empty set $T$ of formulas-very often restricted to formulas obeying certain syntactical restrictions (e.g. only a given finite set of function symbols or predicate symbols may be allowed)- which is closed under consequence. More precisely, $T$ is a theory, if for all $F_1, F_2, \ldots, F_n \in T$ and formulas $G$, if $G$ is a consequence of $F_1, F_2, \ldots, F_n$ then $G \in T$. The formulas which are elements of a theory $T$ are called theorems of $T$''}.

\subsubsection{Elements of a theory}\label{app:a2} As can be seen from the previous section, there is no unanimous agreement on the definition of the concept of theory. However, there is a certain consensus on determining the constituent elements of a theory. In general terms, we can talk about: concepts/constructs, propositions/hypotheses, variables/empirical indicators and boundaries.

In the context of organizational behavior, Bacharach \cite[p. 498]{n5} states: \emph{``The components of a theory include constructs, variables, propositions and hypotheses bounded by assumptions. In more detailed terms a theory may be viewed as a system of constructs and variables in which the constructs are related to each other by propositions and the variables are related to each other by hypotheses''}.

Sandberg \& Alvesson \cite{n8} state that a body of knowledge should meet the following structural elements to be considered a theory: a purpose, it should be directed to a phenomenon, It must offer some form of conceptual order, It must provide some intellectual insights about the phenomenon, it needs to include relevance criteria that can be used to evaluate how effectively it performs its overall purpose, It must have some form of empirical support, and a theory is always constrained by boundary conditions.

Dubin \cite{dubin:1978}, one of the first social scientists to formalize theory construction, identifies the following elements of a theory: units, laws of interaction, boundary, system states, propositions, empirical indicators, and testable hypotheses.

It is interesting to note that all Gregor, Dubin, and Johnson \& Ekstedt indicate that hypotheses (relationships between constructs) must be testable. Specifically, Gregor \cite{gregor:2006}, and Johnson \& Ekstedt \cite{n11} emphasize that theories should be falsifiable. This is a term coined by Popper \cite{n10} who distinguished between empirical sciences (including software engineering \cite{n11} and metaphysical systems, referring to the criterion that separates them as ``The Problem of Demarcation.'' For Popper, the boundary between empirical science and other domains is defined by the principle of falsifiability. This principle asserts that theories are not empirically verifiable but must be falsifiable. A theory formulated through inductive reasoning can never be proven to be ``always true'', as empirical evidence may eventually emerge that refutes it. To illustrate this principle, Popper presents the statement: ``All swans are white''. According to him, this assertion is not scientific in the strict sense because it cannot be definitively verified. No matter how many white swans are observed, there could always exist a swan of another colour. The only way to prove the statement false is to find a black swan, or one of any colours other than white. Therefore, the set of hypotheses (formulas) that constitute a theory must be testable, i.e. verifiable and contrastable, against empirical facts. It is important to note that, as Popper \cite[p. 51]{n10} states: \emph{``the falsification of a logically deduced statement may sometimes not affect the whole system but only some part of it, which may then be regarded as falsified.''} From this, it follows that a theoretical system that is not falsifiable does not belong to the domain of empirical science.

\subsubsection{Types of theory}\label{app:a2-types} Gregor \cite[p. 620]{gregor:2006} establishes the following taxonomy for types of theories in information systems:

\textbf{I. Analysis}. Says what is. The theory does not extend beyond analysis and description. No causal relationships among phenomena are specified and no predictions are made.

\textbf{II. Explanation}. Says what is, how, why, when, and where. The theory provides explanations but does not aim to predict with any precision. There are no testable propositions.

\textbf{III. Prediction}. Says what is and what will be. The theory provides predictions and has testable propositions but does not have well-developed justificatory causal explanations.

\textbf{IV. Explanation and prediction (EP)}. Says what is, how, why, when, where, and what will be. Provides predictions and has both testable propositions and causal explanations.

\textbf{V. Design and action}. Says how to do something. The theory gives explicit prescriptions (e.g., methods, techniques, principles of form and function) for constructing an artefact.

Sandberg \& Alvesson \cite{n8} propose five major types of theory:

\begin{itemize}
\item
  \textbf{Explaining}: This type is used mainly in positivistic approaches to the social sciences, but also frequently within various social-constructionist paradigms. The theory should postulate what variables are part of the explanation, needs to show how these variables are causally related to each other, and it should explain why the relationships exist among the variables. Finally, the theory should express its boundary conditions in the sense of whom the theory applies to, when it applies, and where it applies.
\item
  \textbf{Comprehending}: These theories are typically, but not necessarily, interpretative and qualitative in character. The main purpose of comprehending theory is to offer a qualified understanding of organizational phenomena by determining their meaning: that is, what phenomena such as ``decision,'' ``diversity,'' or ``identity'' are about.
\item
  \textbf{Ordering}: These theories can be both positivistic and interpretive and therefore linked to different social science traditions. The main purpose of ordering theory is to categorize phenomena in theoretically useful ways.
\item
  \textbf{Enacting}: The overall purpose of enacting theory is to articulate how phenomena are continuously produced and reproduced: that is, the processes through which they emerge, evolve, reoccur, change and decline over time.
\item
  \textbf{Provoking}: Its main purpose is to show alternative, often eye-opening and disruptive ways of seeing phenomena.
\end{itemize}

Finally, in the context of software engineering, Ralph \cite{ralph:2019} distinguished between theories that explain why things happen (variance theories), theories that explain how things happen (process theories), and theories to analyze and understand situations (taxonomies).

Note that Gregor's \cite{gregor:2006} ``I. Analysis'' theory type is similar to Sandberg \& Alvesson's \cite{n8} ``Ordering'' theory type and Ralph's \cite{ralph:2019} ``taxonomies'' theory. All of them refer to establishing taxonomies, frameworks, or classification schemes. In this same sense, similarities can be found between the two types ``Explanation'' and between ``Design and action'', ``Enacting'', and ``Process'' theories.

\subsection{Theory building process}\label{theory-building-process-1}

This appendix describes three proposals for developing theories.

\subsubsection{Lynham's general method of theory construction}\label{sec:lynham-general} Lynham \cite{lynham2002} proposes a recursive 5-phase system for theory building:

\begin{itemize}
\item
  Conceptual development includes the development of the core elements of the theory, their interrelationships, and the conditions and limits within which the theory will operate.
\item
  Operationalization includes the conversion of key elements into observable entities in the empirical world, as well as their interrelationships.

  \begin{itemize}
  \item
    Confirmation or disconfirmation. It involves designing and implementing a research agenda that proposes studies to confirm/refute the theory.
  \item
    Application. A confirmed theory is not enough. It needs to be applied to the domain of the problem for which it was proposed. The experience resulting from this application allows us to obtain more information and refine the theory.
  \item
    Continuous refinement and development (of the theory). The theory must adapt to new observations and rectify its hypotheses if they are refuted by new data. It is a process of continuous improvement.
  \end{itemize}
\end{itemize}

\subsubsection{Ralph's process theories}\label{sec:ralph-process} Ralph \cite{ralph:2019} recommends a set of seven steps for generating taxonomies and process theory.

\begin{enumerate}
\def\labelenumi{\arabic{enumi}.}
\item
  Choose a strategy. Select appropriate sources of information. According to the author, the best ones are: secondary studies, interpretive case studies or grounded theory studies.
\item
  Site selection. ``Because the goal here is to generate a theory, not to statistically generalize results, the most important criteria are that the site is accessible and produces rich data. Authors should not need a litany of justifications''
\item
  Data Collection. A typical investigation might include direct observation, interviewing participants and collecting documents and archival data.
\item
  Data Analysis. The author details the following strategies for data analysis: initial coding, in vivo coding, descriptive coding, memoing and theoretical coding, process coding, versus coding.
\item
  Conceptual Evaluation. The author suggests adopting the quality criteria described by Lincoln and Guba \cite{lincoln:1985}: credibility, transferability, dependability and confirmability.
\item
  Writing-Up. The author makes a series of recommendations to help the reader understand the theories developed, including the following: i) explain why the proposed theory is needed; ii) define the elements of the theory clearly and carefully (construct, agent, instance, object, . . . ); iii) include a detailed conceptual evaluation of the proposed theory.
\item
  Peer Review\textbf{.} The author lists 17 recommendations for reviewing the theory, a review that should focus on validity in the broad sense of methodological correctness.
\end{enumerate}

\subsubsection{Rivard's spiral model}\label{sec:rivard-spiral} The author \cite{rivard2021theory} proposes a spiral process with six iterations and two components: activities and products (ions). The activities `read', `reflect', and `write' are carried out in each iteration, although they take different forms depending on the specific iteration. The proposed process emphasizes cohesion throughout the development of all iterations. Cohesion means that all products must be aligned when they form part of a whole; they must be consistent. Table \ref{tab:activities-ions} shows the activities and products of each iteration.

\begin{table}[h]
\centering
\caption{Activities and ions}\label{tab:activities-ions}
\begin{tabular}{lll}
\toprule
iteration & Activities & Products (ion) \\
\midrule
 & read, reflect, write & erudition (1) \\
 & read, reflect, write & motivation (2) \\
 & read, reflect, write & definition (3) \\
 & research, read, reflect, write & imagination (4) \\
 & respite, read, reflect, write & explanation and presentation (5) \\
 & review, renounce, revise, read, reflect, write & contribution (6) \\
\bottomrule
\end{tabular}
\end{table}

\begin{enumerate}
\def\labelenumi{(\arabic{enumi})}
\item
  Erudition refers to the breadth and depth of one's understanding of a particular topic, ``acquired by study, research, etc.'' The objective is to extract the essence of what is known on a phenomenon of interest, to make sense of what others have written, and to map the territory of extant research.
\item
  The objective is to identify gaps in extant explanations or to relax the assumptions that undergird extant research, with the aim of expanding extant knowledge on the phenomenon of interest.
\item
  The phenomenon of interest, the type of theory that is proposed, the boundary of the theory, and theory constructs.
\item
  Generative theorizing practices (analogizing, metaphorizing, mythologizing, and modelling), modelling as theorizing, Applying heuristics. The aim is to generate ideas about the relationships between constructs.
\item
  Explanation refers to causal relationships. The presentation of the theory includes the structure of the manuscript, the syntax, the vocabulary, and sometimes a graphical representation, which are powerful tools that authors can use to convey their novel explanation.
\item
  How their theory is novel and different from extant explanations of the phenomenon of interest; what it can explain that other theoretical explanations have missed or ignored; how their theoretical explanation adds to extant knowledge about a phenomenon, to their field, and to other fields, of study.
\end{enumerate}

Although the process indicates the activities to be carried out and their sequencing, it does not detail, for example, how the constructs and hypotheses that establish relationships among them are generated. As shown in Section \ref{sec:framework}, the methodological framework proposed in this work addresses this common limitation by providing a set of explicit steps that guide the generation of constructs, the specification of their values, and the transformation of propositions into testable hypotheses.

\clearpage

\section{Fundamental notions of propositional logic.}\label{fundamental-notions-of-propositional-logic.}

In this appendix, we shall briefly describe the fundamentals of a theory of propositional logic, also known as propositional calculus or sentential logic, with variables. This is a simple extension regular propositional logic to fit with the purposes of this paper. The interested reader can deepen in this formalism by checking \cite{van1994logic}. This propositional logic will be useful for the discussion developed in Section \ref{operationalization-1}.

\subsection{Propositional logic}\label{propositional-logic}

In standard propositional logic with variables we have an infinite collection of \emph{literals} denoted by $P, P_1, P_2, Q, Q_1, \ldots$ or similar symbols. A \emph{formula}, also known as a \emph{hypothesis} in the context of qualitative research, in propositional logic is a string of symbols generated recursively by the following rules:
\begin{itemize}
    \item A literal is a formula.
    \item If $\varphi$ is a formula, then $\neg \varphi$ is a formula. The new formula must be understood as ``not $\varphi$''.
    \item If $\varphi$ and $\psi$ are formulas, then $\varphi \vee \psi$ is a formula. The new formula must be understood as ``$\varphi$ or $\psi$''.
    \item If $\varphi$ and $\psi$ are formulas, then $\varphi \wedge \psi$ is a formula. The new formula must be understood as ``$\varphi$ and $\psi$''.
    \item If $\varphi$ and $\psi$ are formulas, then $\varphi \to \psi$ is a formula. The new formula must be understood as ``$\varphi$ implies $\psi$''.
    \item If $\varphi$ and $\psi$ are formulas, then $\varphi \leftrightarrow \psi$ is a formula. The new formula must be understood as ``$\varphi$ if and only if $\psi$''.
\end{itemize}

Some examples of formulas are $\varphi = P$, $\varphi = P \wedge (\neg Q)$, or $\varphi = (\neg P) \to (P \wedge Q)$. The collection of all formulas will be denoted by $\textbf{For}$.

In this context, a \emph{truth assignment} is a function
$$
	\nu: \textbf{For} \to \left\{0,1\right\}
$$
such that
\begin{align*}
		\nu(\psi \wedge \varphi) & = \left\{\begin{matrix}1 & \textrm{if } \nu(\psi) = 1 \textrm{ and } \nu(\varphi) = 1,\\
		0 & \textrm{if } \nu(\psi) = 0 \textrm{ or } \nu(\varphi) = 0, \end{matrix}\right.\\
		\nu(\psi \vee \varphi) & = \left\{\begin{matrix}1 & \textrm{if } \nu(\psi) = 1 \textrm{ or } \nu(\varphi) = 1,\\
		0 & \textrm{if } \nu(\psi) = 0 \textrm{ and } \nu(\varphi) = 0, \end{matrix}\right.\\
		\nu(\psi \to \varphi) &=  \left\{\begin{matrix}1 & \textrm{if } \nu(\psi) = 0 \textrm{ or } (\nu(\psi) = 1 \textrm{ and } \nu(\varphi) = 1),\\
		0 & \textrm{if } \nu(\psi) = 1 \textrm{ and } \nu(\varphi) = 0, \end{matrix}\right.\\
		\nu(\psi \leftrightarrow \varphi) &= \left\{\begin{matrix}1 & \textrm{if } \nu(\psi) = \nu(\varphi), \\
		0 & \textrm{if } \nu(\psi) \neq \nu(\varphi), \end{matrix}\right.\\
		\nu(\neg \psi) &= 1-\nu(\psi).
\end{align*}
If a formula $\varphi$ satisfies $\nu(\varphi) = 1$, we say that $\varphi$ is \emph{true} under the interpretation $\nu$, and \emph{false} otherwise.
Notice that, due to the construction rules for formulas, a truth assignment is fully determined by the truth values $\nu(P)$ for each literal $P \in \mathcal{P}$.

A collection $T \subseteq \textbf{For}$ of formulas is called a \emph{theory}. A formula $\varphi \in \textbf{For}$ is said to be a \emph{consequence} of the theory $T$ if for all truth assignments $\nu$ such that $\nu(\psi) = 1$ for all $\psi \in T$ (i.e.\ all truth assignments that make $T$ true), we also have $\nu(\varphi) = 1$. We denoted this consequence $T \models \varphi$. In particular, if $T$ is empty, then $\nu(\varphi) = 1$ for all truth assignments $\nu$, and then we say that $\varphi$ is a \emph{tautology}, denoted by $\models \varphi$. Observe that if $T$ is finite, say $T = \{\psi_1, \ldots, \psi_n\}$, then $T \models \varphi$ is equivalent to the fact that $\psi_1 \wedge \ldots \wedge \psi_n \to \varphi$ is a tautology, i.e.\ $\models (\psi_1 \wedge \ldots \wedge \psi_n \to \varphi)$. On the contrary, a theory $T$ is said to be \textit{unsatisfiable} if for every truth assignment $\nu$, there exists $\psi \in T$ such that $\nu(\psi) = 0$.

For example, let us consider the theory $T = \{P, P\to Q\}$. Then, for the formula $\varphi = Q$ we have that $T \models Q$, since $\left(P \wedge (P \to Q)\right) \to Q$ is a tautology. Another example of tautology is the `principle of the excluded middle' $\models (P \vee \neg P)$ for each literal $P$. 

\subsection{Propositional logic with variables}\label{propositional-logic-with-variables}

Despite the power of propositional logic, for this work we will need to consider a simple extension of the propositional logic by adding variables. This will add an extra layer of complexity providing a logic in the midpoint between regular propositional logic and first-order logic that will be very useful for our purposes. Roughly speaking, the idea is that in this case the literals are no longer atomic, but composed of more complex sentences.

In this manner, this logic is defined on a certain \emph{language}, which is a tuple $\mathcal{L} = (\mathcal{U}, \mathcal{V}, \mathcal{F}, \mathcal{R})$ of:
\begin{itemize}
    \item A finite set of \emph{universes} $\mathcal{U} = (U_1, U_2, \ldots, U_n)$, that represent the possible types of the variables considered.
    \item A set of \emph{variables} $\mathcal{V} = (V_1, V_2, \ldots, V_n)$, that represent values that will be taken when interpreted. The variables of $V_i$ are thought to have type in the universe $U_i$.
    \item A set of \emph{functions} $\mathcal{F} = (F_1, F_2, \ldots, F_n)$, that represent functions when interpreted, one collection of functions per universe. Each function will have an associated \emph{arity}, which is a positive natural number representing its number of arguments.
    \item A set of \emph{relations} $\mathcal{R} = (R_1, R_2, \ldots, R_n)$, that will represent relations to be checked when interpreted. Each relation will have an associated \emph{arity} which is a non-negative number.
\end{itemize}

In this setting, we can define \emph{terms} with types recursively as follows:
\begin{itemize}
    \item For each $i$, a variable $x \in V_i$ is a term of type $U_i$.
    \item For each $i$, a value $c \in U_i$ is a term of type $U_i$.
    \item If $t_1, \ldots, t_n$ are terms of type $U_i$ (same $i$ for all of them), and $f \in F_i$ is a function of arity $n$, then $f(t_1, \ldots, t_n)$ is a term of type $U_i$.
\end{itemize}
For example, in a language with a unique universe $\mathcal{U} = (U_1 = \mathbb{R})$ of real numbers, $\mathcal{V} = (V_1 = \{a, b\})$ and functions $\mathcal{F} = (F_1 = \{\sin, +\})$, with $\sin$ is an unary function and $+$ a binary function, the following are terms of type $U_1$: $a$, $\sin(a)$, $a + b$ and $\sin(a) + b$.

With these terms, a \emph{literal} is given by a string of the form $R(t_1, \ldots, t_n)$, where $R \in R_i$ is a relation of arity $n\geq 0$ and $t_1, \ldots, t_n$ are terms of type $U_i$. Following the previous example, if we consider the relations $\mathcal{R} = (R_1 = \{\leq, \neq 0\})$ with $\leq$ being a binary relation and $\neq 0$ a unary relation, then $\sin(a) \leq b$, $\sin(a) + b \leq a + b$ or $\sin(a) \neq 0$ are literals of type $U_1$. Notice that a relation of arity $0$ should be interpreted as a literal with no variables, as in the previous case of regular propositional logic.

From these literals, the formulas are constructed in the same way as in the standard propositional logic, by composing literals with the operators $\wedge, \vee, \neg, \to$ and $\leftrightarrow$. The set of formulas constructed in this way is denoted by $\textbf{For}$.

In this logic, a truth assignment is induced by a \emph{model}. A model $\mathcal{M}$ is an assignment that converts every variable $x \in V_i$ of type $U_i$ into a value $c^{\mathcal{M}} \in U_i$. Analogously, every term $t$ of type $U_i$ is converted into a value $t^{\mathcal{M}} \in U_i$. In this manner, a model induces a truth assignment $\nu_{\mathcal{M}}: \textbf{For} \to \{0,1\}$ by setting that, for a literal of the form $R(t_1, \ldots, t_n)$ with $R$ a relation and $t_1, \ldots, t_n$ terms of type $U_i$, we have $\nu_{\mathcal{M}}(R(t_1, \ldots, t_n)) = 1$ if $R(t_1^{\mathcal{M}}, \ldots, t_n^{\mathcal{M}})$ holds as a relation in $U_i$.

With these ideas, we obtain analogous notions of theory $T \subseteq \mathbf{For}$ and consequence $T \models \varphi$ if for every model $\mathcal{M}$ such that $\nu_{\mathcal{M}}(\psi) = 1$ for all $\psi \in T$, we also have $\nu_{\mathcal{M}}(\varphi) = 1$. In particular, a tautology is a formula $\varphi$ such that $\nu_{\mathcal{M}}(\varphi) = 1$ for every model $\mathcal{M}$. Analogously, a theory $T$ is said to be unsatisfiable if for every model $\mathcal{M}$, there exists $\psi \in T$ such that $\nu_{\mathcal{M}}(\psi) = 0$.

\subsection{Automatic deduction}\label{automatic-deduction}

In this section, we shall briefly review some of the existing methods for automatic deduction in propositional logic. For a more thorough introduction, see \cite{harrison2009handbook}. We will focus on propositional calculus since, for the purposes of deduction, we can consider as `atomic' the literals in the propositional calculus with variables, so that a literal of the form $R(t_1, \ldots, t_n)$ is just treated as a literal $R$. 

Suppose that $T$ is a theory and $\varphi \in \textbf{For}$ is a formula. Observe that $\varphi$ is a consequence of $T$, that is $T \models \varphi$, if and only if $T \cup \{\neg \varphi\}$ is an unsatisfiable theory. In this manner, the most effective methods for automatic deduction are based exactly on unsatisfiability detectors.

The most common method for checking this unsatisfiability is the so-called Davis-Putnam algorithm. It is based on the resolution rule, that claims that, for any formulas $\psi, \varphi \in \textbf{For}$ and any literal $P$, we have
$$
	\{\psi \vee P, \varphi \vee \neg P\} \models \psi \vee \varphi.
$$

To apply it, recall that any formula $\varphi \in \textbf{For}$ can be put in conjuctive normal form (CNF) as an equivalent formula that is a juxtaposition of and's of or's, that is
$$
    \varphi \equiv C_1 \wedge \ldots \wedge C_n,
$$
where $C_i = \ell_{i,1} \vee \ldots \vee \ell_{i, m_i}$ are the so-called clauses, with $\ell_{i,j}$ literals or negation of literals. For instance, the CNF of the formula $P \to (Q \wedge R)$ is $(\neg P \vee Q) \wedge (\neg P \vee R)$. In this manner, given a theory $T$, we can form its \emph{clausular theory} $T_c$ as the equivalent theory formed by the clauses of the CNFs of the formulas in $T$. Using the fact that a clausular theory is unsatisfiable if and only if the empty clause can be obtained by applying the resolution rule reiteratively, we obtain the Davis-Putnam algorithm, as described in Algorithm \ref{alg:davis-putnam}.

\begin{algorithm}[H]
 \KwData{Theory $T$}
 \KwResult{Whether $T$ is unsatisfiable or not}
 $X := $ Clausular form of $T$\;
 \While{\textbf{not} $\emptyset \in X$}{
  $\mathcal{R}(X) := $ Set of all the possible resolutions in $X$\;
  \If{$\mathcal{R}(X) = X$}{
   \textbf{return} $T$ is satisfiable\;
   }{
   $X := \mathcal{R}(X)$\;
  }
 }
 \textbf{return} $T$ is unsatisfiable\;
 \caption{Davis-Putnam algorithm}
 \label{alg:davis-putnam}
\end{algorithm}

It is worth mentioning that, in the case that all the clauses in $T_c$ are Horn clauses, i.e.\ clauses with at most one non-negated literal, then the automatic deduction can be performed even more efficiently by using the so-called SLD (Selected Linear Definite) algorithm, whose pseudo-code can be checked in \cite{Davis-PutnamAlg,kowalski1971linear}.

With this tool at hand to decide whether $T \models \varphi$ for any formula $\varphi$, in the case that $T$ is finite, we can extract a `minimal equivalent theory' $T_0 \subseteq T$ from $T$. This theory satisfies that $T$ is equivalent to $T_0$ and no formula can be removed from $T_0$ without losing the equivalence with $T$. 

\begin{algorithm}[H]
 \KwData{Theory $T$}
 \KwResult{A minimal equivalent theory $T_0 \subseteq T$}
 $T_0 := T$\;
 \For{$\varphi \in T_0$}{
  \If{$T_0 - \{\varphi\} \models \varphi$ (using Algorithm \ref{alg:davis-putnam})}{
   $T_0 := T_0 - \{\varphi\}$\;
   }{
  }
 }
 \textbf{return} $T_0$\;
 \caption{Computation of a minimal theory}
 \label{alg:minimal-theory}
\end{algorithm}

Observe that the minimal theory $T_0$ is not unique, and the result depends on the order in which the formulas are checked at the `for' statement in Algorithm \ref{alg:minimal-theory}.

\subsection{Automatic deduction in implication theories}\label{automatic-deduction-in-implication-theories}

In the case that the theory $T$ is simple, we can perform automatic deduction even more easily than with Algorithm \ref{alg:davis-putnam} by using some graph theory methods. In this section, we will suppose that our theory $T$ is an \emph{implication theory}, meaning that $T$ is finite and all the formulas $\varphi$ of $T$ are of the form
$$
    \varphi = \ell \to \ell', 
$$
where both $\ell$ and $\ell'$ are either literals or negation of literals.

In this case, let $P_1, \ldots, P_n$ be all the literals appearing in $T$. We can represent $T$ as a graph $\mathcal{G}_T$ whose vertices are labelled with $P_1, \ldots, P_n$ and its negations $\neg P_1, \ldots, \neg P_n$. For each formula of the form $\ell \to \ell'$ in $T$, we draw a directed edge between the vertex $\ell$ and the vertex $\ell'$ as well as another one between $\neg \ell'$ and $\neg \ell$ (understanding that $\neg\neg P = P$ for a literal $P$).

\begin{ex}\label{ex:graph-theory}
To illustrate this idea, suppose that we have a theory comprised of the following formulas
$$
    T = \left\{Q \to R, P \to R, S \to Q, \neg S \to \neg P\right\}.
$$
In that case, the associated graph has vertices $P, Q, R, S, \neg P, \neg Q, \neg R$ and $\neg S$ with the edges shown in Figure \ref{fig:graph-example}.

\begin{center}
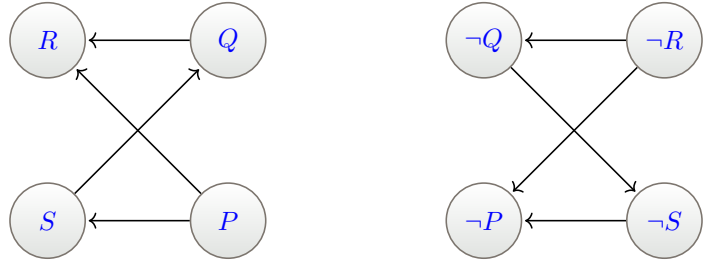

 \begin {tikzpicture}[-latex , shorten >=1pt, auto ,node distance =2.4 cm and 2cm,
semithick ,
state/.style ={ circle ,top color =white , bottom color = processblue!20 ,
draw,processblue , text=blue , minimum width =1 cm}]
   \node[state] (P) {$P$}; 
   \node[state] (Q) [above of=P] {$Q$}; 
   \node[state] (R) [left of=Q] {$R$}; 
   \node[state] (S) [left of=P] {$S$}; 
   \node[state] (nP) [right of=P, xshift = 1cm] {$\neg P$}; 
   \node[state] (nQ) [above of=nP] {$\neg Q$}; 
   \node[state] (nR) [right of=nQ] {$\neg R$}; 
   \node[state] (nS) [right of=nP] {$\neg S$}; 

    \path[->] 
        (S) edge node {} (Q)
        (Q) edge node {} (R)
        (P) edge node {} (R)
        (P) edge node {} (S)

        (nQ) edge node {} (nS)
        (nR) edge node {} (nQ)
        (nR) edge node {} (nP)
        (nS) edge node {} (nP);
\end{tikzpicture}
    \captionof{figure}{Example of the graph associated with an implication theory.}
    \label{fig:graph-example}
\end{center}
\end{ex}

From this graph $\mathcal{G}_T$ representing the theory, we exploit its structure to infer properties of the theory $T$ by simply applying graph algorithms. An interesting information is the \emph{transitive closure} of $\mathcal{G}_T$, which is a graph $\overline{\mathcal{G}}_T$ with the same vertices as $\mathcal{G}_T$ and such that an edge $\ell \to \ell'$ exists in $\overline{\mathcal{G}}_T$ if and only if there exists a directed path from $\ell$ to $\ell'$ in $\mathcal{G}_T$. Equivalently, this means that $\ell'$ is inferable from $\ell$ using the theory $T$ by combining several of its implications, so that the edges in $\overline{\mathcal{G}}_T$ represent all the possible implications derived from $T$.

There exist several ways of computing the transitive closure of $\mathcal{G}_T$. The most direct one is to compute the adjacency matrix $A_{\mathcal{G}_T}$ and to calculate the products $A_{\mathcal{G}_T}^k$, for $1 \leq k \leq n$, where $n$ is the number of vertices in $\mathcal{G}_T$. This matrix $A_{\mathcal{G}_T}^k$ contains a non-zero entry in the position $(i,j)$ if and only if there exists a directed path starting in the $i$-th vertex and ending in the $j$-th vertex with length exactly $k$. With this information, we can form
$$
    B_{\mathcal{G}_T} = \sum_{k=1}^n A_{\mathcal{G}_T}^k,
$$
which represents all the vertices reacheable in $\mathcal{G}_T$. Therefore, the adjacency matrix of $\overline{\mathcal{G}}_T$ is precisely the binarization of $B_{\mathcal{G}_T}$, i.e.\ the $(i,j)$-entry of $A_{\overline{\mathcal{G}}_T}$ is $1$ if the $(i,j)$-entry of $B_{\mathcal{G}_T}$ is positive, and it is $0$ otherwise. However, since matrix multiplication might be expensive computationally, alternative methods can be used, such as the Floyd–Warshall algorithm \cite{cormen2022introduction} for minimum distance in a graph, so that again the edges in $\overline{\mathcal{G}}_T$ are precisely those connecting edges in $\mathcal{G}_T$ at a finite distance.

\begin{ex}
    Coming back to our Example \ref{ex:graph-theory}, the adjacency matrix of $\mathcal{G}_T$ is the following
    $$\tiny
        A_{\mathcal{G}_T} = \begin{pmatrix}
            0 & 0 & 1 & 1 & 0 & 0 & 0 & 0 \\
            0 & 0 & 1 & 0 & 0 & 0 & 0 & 0 \\
            0 & 0 & 0 & 0 & 0 & 0 & 0 & 0 \\
            0 & 1 & 0 & 0 & 0 & 0 & 0 & 0 \\
            0 & 0 & 0 & 0 & 0 & 0 & 0 & 0 \\
            0 & 0 & 0 & 0 & 0 & 0 & 0 & 1 \\
            0 & 0 & 0 & 0 & 1 & 1 & 0 & 0 \\
            0 & 0 & 0 & 0 & 1 & 0 & 0 & 0 \\
        \end{pmatrix}.
    $$
In this manner, if we compute $B_{\mathcal{G}_T} = A_{\mathcal{G}_T} + A_{\mathcal{G}_T}^2 + \ldots + A_{\mathcal{G}_T}^8$, we get
$$\tiny
    B_{\mathcal{G}_T} = \left(\begin{array}{rrrrrrrr}
0 & 1 & 2 & 1 & 0 & 0 & 0 & 0 \\
0 & 0 & 1 & 0 & 0 & 0 & 0 & 0 \\
0 & 0 & 0 & 0 & 0 & 0 & 0 & 0 \\
0 & 1 & 1 & 0 & 0 & 0 & 0 & 0 \\
0 & 0 & 0 & 0 & 0 & 0 & 0 & 0 \\
0 & 0 & 0 & 0 & 1 & 0 & 0 & 1 \\
0 & 0 & 0 & 0 & 2 & 1 & 0 & 1 \\
0 & 0 & 0 & 0 & 1 & 0 & 0 & 0
\end{array}\right).
$$
Therefore, the adjacency matrix of the transitive closure graph is
$$\tiny
    A_{\overline{\mathcal{G}}_T} = \left(\begin{array}{rrrrrrrr}
0 & 1 & 1 & 1 & 0 & 0 & 0 & 0 \\
0 & 0 & 1 & 0 & 0 & 0 & 0 & 0 \\
0 & 0 & 0 & 0 & 0 & 0 & 0 & 0 \\
0 & 1 & 1 & 0 & 0 & 0 & 0 & 0 \\
0 & 0 & 0 & 0 & 0 & 0 & 0 & 0 \\
0 & 0 & 0 & 0 & 1 & 0 & 0 & 1 \\
0 & 0 & 0 & 0 & 1 & 1 & 0 & 1 \\
0 & 0 & 0 & 0 & 1 & 0 & 0 & 0
\end{array}\right),
$$
whose associated graph is shown in Figure \ref{fig:graph-example-transitive-closure}.

\begin{center}
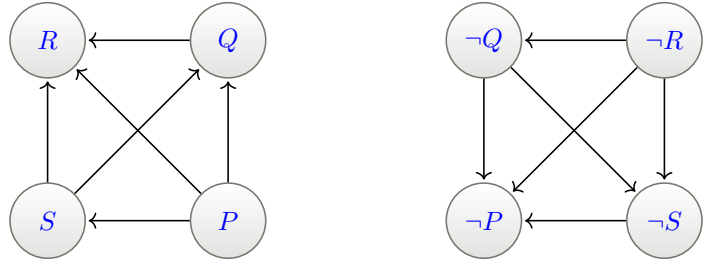

 \begin {tikzpicture}[-latex , shorten >=1pt, auto ,node distance =2.4 cm and 2cm,
semithick ,
state/.style ={ circle ,top color =white , bottom color = processblue!20 ,
draw,processblue , text=blue , minimum width =1 cm}]
   \node[state] (P) {$P$}; 
   \node[state] (Q) [above of=P] {$Q$}; 
   \node[state] (R) [left of=Q] {$R$}; 
   \node[state] (S) [left of=P] {$S$}; 
   \node[state] (nP) [right of=P, xshift = 1cm] {$\neg P$}; 
   \node[state] (nQ) [above of=nP] {$\neg Q$}; 
   \node[state] (nR) [right of=nQ] {$\neg R$}; 
   \node[state] (nS) [right of=nP] {$\neg S$}; 

    \path[->] 
        (S) edge node {} (R)
        (S) edge node {} (Q)
        (Q) edge node {} (R)
        (P) edge node {} (Q)
        (P) edge node {} (R)
        (P) edge node {} (S)

        (nQ) edge node {} (nP)
        (nQ) edge node {} (nS)
        (nR) edge node {} (nQ)
        (nR) edge node {} (nP)
        (nR) edge node {} (nS)
        (nS) edge node {} (nP);
\end{tikzpicture}
    \captionof{figure}{Transitive closure of the graph in Figure \ref{fig:graph-example}.}
    \label{fig:graph-example-transitive-closure}
\end{center}
\end{ex}

It is interesting to notice that, once we know the transitive closure of $\mathcal{G}_T$, it is possible to extract a minimal equivalent subgraph with the fewest connections. Explicitly, a \emph{transitive reduction} of a graph $\mathcal{G}_T$ is a subgraph $\mathcal{G}_T^0$ with the same vertices as $\mathcal{G}_T$ and the fewest possible edges such that its transitive closure coincides with the one of $T$ i.e.\ $\overline{\mathcal{G}}_T^0 = \overline{\mathcal{G}}_T$. A transitive reduction of an acyclic graph $\mathcal{G}_T$ can be easily computed as the graph whose adjacency matrix is the binarization of $A_{\mathcal{G}_T} - A_{\mathcal{G}_T}\cdot A_{\overline{\mathcal{G}}_T} = A_{\mathcal{G}_T} (\textup{Id} - A_{\overline{\mathcal{G}}_T})$. Indeed, the product $A_{\mathcal{G}_T} \cdot A_{\overline{\mathcal{G}}_T}$ represents those vertices that are connected by a path of length greater or equal than two so, if we subtract them from $A_{\mathcal{G}_T}$, we exactly get the irredundant edges.

\begin{ex}
    Continuing with Example \ref{ex:graph-theory}, the product $A_{\mathcal{G}_T} \cdot A_{\overline{\mathcal{G}}_T}$ is
    $$\tiny
        A_{\mathcal{G}_T} \cdot A_{\overline{\mathcal{G}}_T} = \left(\begin{array}{rrrrrrrr}
0 & 1 & 1 & 0 & 0 & 0 & 0 & 0 \\
0 & 0 & 0 & 0 & 0 & 0 & 0 & 0 \\
0 & 0 & 0 & 0 & 0 & 0 & 0 & 0 \\
0 & 0 & 1 & 0 & 0 & 0 & 0 & 0 \\
0 & 0 & 0 & 0 & 0 & 0 & 0 & 0 \\
0 & 0 & 0 & 0 & 1 & 0 & 0 & 0 \\
0 & 0 & 0 & 0 & 1 & 0 & 0 & 1 \\
0 & 0 & 0 & 0 & 0 & 0 & 0 & 0
\end{array}\right).
    $$
Therefore, we have that
$$\tiny
    A_{\mathcal{G}_T} - A_{\mathcal{G}_T} \cdot A_{\overline{\mathcal{G}}_T} = \left(\begin{array}{rrrrrrrr}
0 & -1 & 0 & 1 & 0 & 0 & 0 & 0 \\
0 & 0 & 1 & 0 & 0 & 0 & 0 & 0 \\
0 & 0 & 0 & 0 & 0 & 0 & 0 & 0 \\
0 & 1 & -1 & 0 & 0 & 0 & 0 & 0 \\
0 & 0 & 0 & 0 & 0 & 0 & 0 & 0 \\
0 & 0 & 0 & 0 & -1 & 0 & 0 & 1 \\
0 & 0 & 0 & 0 & 0 & 1 & 0 & -1 \\
0 & 0 & 0 & 0 & 1 & 0 & 0 & 0
\end{array}\right).
$$
Therefore, the adjacency matrix of the transitive closure graph is the binarization of the previous matrix, which is
$$\tiny
    A_{{\mathcal{G}}_T^0} = \left(\begin{array}{rrrrrrrr}
0 & 0 & 0 & 1 & 0 & 0 & 0 & 0 \\
0 & 0 & 1 & 0 & 0 & 0 & 0 & 0 \\
0 & 0 & 0 & 0 & 0 & 0 & 0 & 0 \\
0 & 1 & 0 & 0 & 0 & 0 & 0 & 0 \\
0 & 0 & 0 & 0 & 0 & 0 & 0 & 0 \\
0 & 0 & 0 & 0 & 0 & 0 & 0 & 1 \\
0 & 0 & 0 & 0 & 0 & 1 & 0 & 0 \\
0 & 0 & 0 & 0 & 1 & 0 & 0 & 0
\end{array}\right).
$$
The associated graph is shown in Figure \ref{fig:transitive-reduction}. We can observe that this subgraph contains no redundant links for deduction and that $\overline{\mathcal{G}_T^0} =\overline{\mathcal{G}}_T$.

\begin{center}
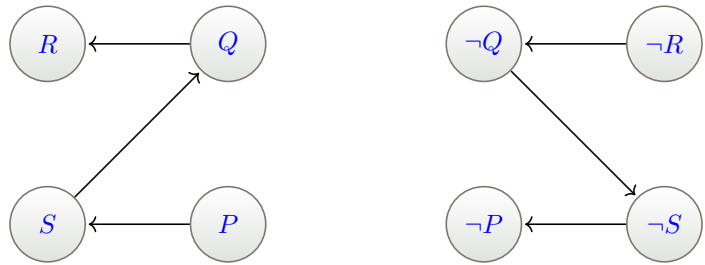

 \begin {tikzpicture}[-latex , shorten >=1pt, auto ,node distance =2.4 cm and 2cm,
semithick ,
state/.style ={ circle ,top color =white , bottom color = processblue!20 ,
draw,processblue , text=blue , minimum width =1 cm}]
   \node[state] (P) {$P$}; 
   \node[state] (Q) [above of=P] {$Q$}; 
   \node[state] (R) [left of=Q] {$R$}; 
   \node[state] (S) [left of=P] {$S$}; 
   \node[state] (nP) [right of=P, xshift = 1cm] {$\neg P$}; 
   \node[state] (nQ) [above of=nP] {$\neg Q$}; 
   \node[state] (nR) [right of=nQ] {$\neg R$}; 
   \node[state] (nS) [right of=nP] {$\neg S$}; 

    \path[->] 
        (S) edge node {} (Q)
        (Q) edge node {} (R)
        (P) edge node {} (S)

        (nQ) edge node {} (nS)
        (nR) edge node {} (nQ)
        (nS) edge node {} (nP);
\end{tikzpicture}
    \captionof{figure}{Transitive reduction of the graph in Figure \ref{fig:graph-example}.}
    \label{fig:transitive-reduction}
\end{center}
\end{ex}

If the original graph contains cycles, we first compute the so-called \emph{condensation graph} $\mathcal{G}_T^c$, whose vertices represent strongly connected components of $\mathcal{G}_T$ and an edge between two connected components in $\mathcal{G}_T^c$ exists if and only if there is an edge connecting any two vertices in the corresponding components in the original graph. The condensation graph $\mathcal{G}_T^c$ is acyclic by definition, so we can compute its transitive reduction $(\mathcal{G}_T^c)^0$ using the previous algorithm for acyclic graphs. From this information, the transitive reduction of $\mathcal{G}_T$ can be computed by expanding $(\mathcal{G}_T^c)^0$ by replacing each vertex of the condensation by a cycle with all the vertices in the corresponding strongly connected component of $\mathcal{G}_T$. All the incoming edges to a condensed vertex are connected to one of the vertices of the substituting cycle (arbitrarily, since all are equivalent). For further details, see \cite{aho1972transitive}.

\subsection{Counter-reciprocals}\label{counter-reciprocals}

The counter-reciprocal of an implication such as: $A \to B$ is $\neg B \to \neg A$. 
But, why is it necessary to introduce counter-reciprocals? According to propositional logic, an implication and its counter-reciprocal are logically equivalent:
$$
A \to B \textrm{ is logically equivalent to } \neg B \to \neg A.
$$
This means that both expressions always have the same truth value. Therefore, if hypothesis $A\to B$ is considered, $\neg B \to \neg A$ is also implicitly accepted. On the other hand, the transitivity of implication establishes that
$\left[(A \to B) \wedge (B \to C)\right] \to (A \to C)$.
However, transitivity cannot deduce certain additional implications unless all equivalences of the hypotheses are considered, including counter-reciprocal. Consider the following case in which a theory has the following two formulas:
\begin{equation}\label{eq:1}
  p \to \neg q
\end{equation}
\begin{equation}\label{eq:2}
 r \to q 
\end{equation}
With only these two formulas, it is not possible to deduce that:
	$p \to \neg r$, nor that $r \to \neg p$, which is its counter-reciprocal. 
However, if we add their counter-reciprocal to the two original formulas:
\begin{equation}\label{eq:3}
  q \to \neg p
\end{equation}
\begin{equation}\label{eq:4}
 \neg q \to \neg r
\end{equation}
Now, $p \to \neg r$ can be deduced by the transitive property between (\ref{eq:1}) and (\ref{eq:4}), and $r \to \neg p$ can be deduced by the transitive property between (\ref{eq:2}) and (\ref{eq:3}).

\clearpage

\section{SynT: a software tool for automating the theory synthesis process}\label{synt-a-software-tool-for-automating-the-theory-synthesis-process}

\begin{figure}[h]
  \centering
  \includegraphics[width=0.6\linewidth]{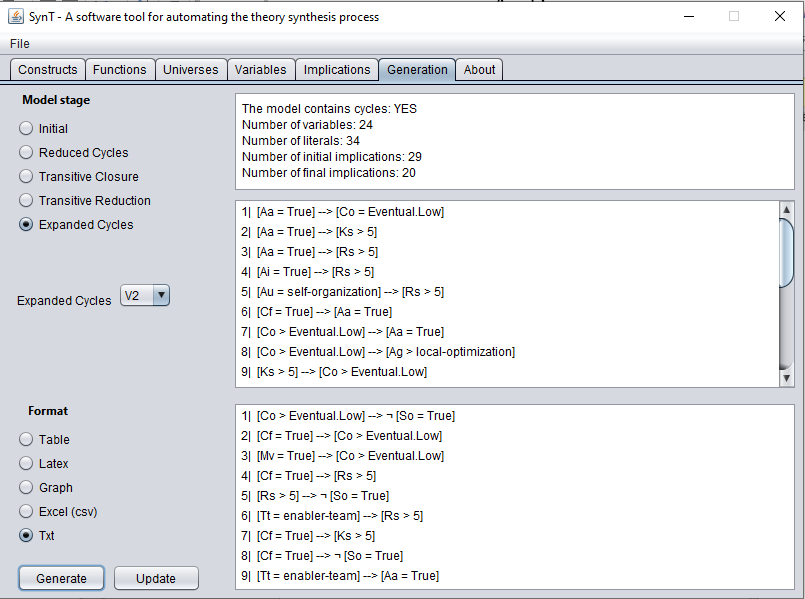}
  \caption{SynT interface.}
  \label{fig:synt-interface}
\end{figure}

SynT is the software tool developed in this research with the aim of reducing the number of hypotheses generated during the operationalization phase by eliminating those hypotheses that can be logically deduced from others. For the use case (DevOps teams) used during the conceptualization and operationalization phases, the tool reduced the number of hypotheses generated during operationalization by 31\%, from the original 29 hypotheses to the final 20 hypotheses. In addition, the tool indicates the chain of deduction that demonstrates how the eliminated hypotheses are derived from the combination of existing ones. Figure \ref{fig:synt-interface} shows the tool interface once the file containing the language developed for the use case (Section \ref{operationalizing-theoretical-concepts-into-constructs}) and the set of defined hypotheses (Appendix \ref{app:d5-hypotheses}) has been loaded.

As can be seen, there are tabs for entering constructs, functions, universes, variables and implications, and a Generation tab that allows you to obtain the final product of the synthesis process or any of the sub-products generated by each of the sub-phases involved in this process. Figure~\ref{fig:synt-interface} shows a final product (Expanded Cycles) and the generation of a text file containing the result of the synthesis process. The panels on the right show data on the initial and final implications (top panel), the set of resulting hypotheses (middle panel) and the set of eliminated hypotheses (bottom panel). As an example, the content of the text file resulting from the synthesis is shown below. Notice that the value ``(Eventual, Low)'' has been codified as ``Eventual.Low'' to prevent parenthesis misinterpretation.

\texttt{--- Implications ---}
\begin{enumerate}
  \item \texttt{{[}Aa = $\mathsf{true}${]} $\to$ {[}Co = Eventual.Low{]}}
  \item \texttt{{[}Aa = $\mathsf{true}${]} $\to$ {[}Ks $>$ 5{]}}
  \item \texttt{{[}Aa = $\mathsf{true}${]} $\to$ {[}Rs $>$ 5{]}}
  \item \texttt{{[}Ai = $\mathsf{true}${]} $\to$ {[}Rs $>$ 5{]}}
  \item \texttt{{[}Au = {self-organization}{]} $\to$ {[}Rs $>$ 5{]}}
  \item \texttt{{[}Cf = $\mathsf{true}${]} $\to$ {[}Aa = $\mathsf{true}${]}}
  \item \texttt{{[}Co $>$ Eventual.Low{]} $\to$ {[}Aa = $\mathsf{true}${]}}
  \item \texttt{{[}Co $>$ Eventual.Low{]} $\to$ {[}Ag $>$ local-optimization{]}}
  \item \texttt{{[}Ks $>$ 5{]} $\to$ {[}Co $>$ Eventual.Low{]}}
  \item \texttt{{[}Mv = $\mathsf{true}${]} $\to$ {[}Aa = $\mathsf{true}${]}}
  \item \texttt{{[}Rd = $\mathsf{true}${]} $\to$ $\neg$ {[}Co $>$ Eventual.Low{]}}
  \item \texttt{{[}Rs $>$ 5{]} $\to$ {[}Cm = $\mathsf{true}${]}}
  \item \texttt{{[}Rs $>$ 5{]} $\to$ {[}Co $>$ Eventual.Low{]}}
  \item \texttt{{[}So = $\mathsf{true}${]} $\to$ $\neg$ {[}Aa = $\mathsf{true}${]}}
  \item \texttt{{[}Tt = {bridge-team}{]} $\to$ {[}So = $\mathsf{true}${]}}
  \item \texttt{{[}Tt = {enabler-team}{]} $\to$ {[}Au = {self-organization}{]}}
  \item \texttt{{[}Tt = {enabler-team}{]} $\to$ {[}Cf = $\mathsf{true}${]}}
  \item \texttt{{[}Tt = {enabler-team}{]} $\to$ {[}Ps = $\mathsf{true}${]}}
  \item \texttt{{[}Tw = $\mathsf{true}${]} $\to$ $\neg$ {[}Rs $>$ 5{]}}
  \item \texttt{$\neg$ {[}Rs $>$ 5{]} $\to$ {[}Tw = $\mathsf{true}${]}}
\end{enumerate}

\texttt{--- Deleted implications ---}
\begin{enumerate}
  \item \texttt{{[}Co $>$ Eventual.Low{]} $\to$ $\neg$ {[}So = $\mathsf{true}${]}}\\
  \noindent Deleted because: \texttt{({[}Co $>$ Eventual.Low{]} $\to$ {[}Aa = $\mathsf{true}${]} $\to$ $\neg$ {[}So = $\mathsf{true}${]})}

  \item \texttt{{[}Cf = $\mathsf{true}${]} $\to$ {[}Co $>$ Eventual.Low{]}}\\

  \noindent Deleted because: \texttt{({[}Cf = $\mathsf{true}${]} $\to$ {[}Aa = $\mathsf{true}${]} $\to$ {[}Ks $>$ 5{]}}
  
  \texttt{$\to$ {[}Co $>$ Eventual.Low{]})}

  \item \texttt{{[}Mv = $\mathsf{true}${]} $\to$ {[}Co $>$ Eventual.Low{]}}\\
  \noindent Deleted because: \texttt{({[}Mv = $\mathsf{true}${]} $\to$ {[}Aa = $\mathsf{true}${]} $\to$ {[}Ks $>$ 5{]} } 
  
  \texttt{$\to$ {[}Co $>$ Eventual.Low{]})}

  \item \texttt{{[}Cf = $\mathsf{true}${]} $\to$ {[}Rs $>$ 5{]}}\\
  \noindent Deleted because: \texttt{({[}Cf = $\mathsf{true}${]} $\to$ {[}Aa = $\mathsf{true}${]} $\to$ {[}Rs $>$ 5{]})}

  \item \texttt{{[}Rs $>$ 5{]} $\to$ $\neg$ {[}So = $\mathsf{true}${]}}\\
  \noindent Deleted because: \texttt{({[}Rs $>$ 5{]} $\to$ {[}Co $>$ Eventual.Low{]} $\to$ {[}Aa = $\mathsf{true}${]} $\to$ $\neg$ {[}So = $\mathsf{true}${]})}

  \item \texttt{{[}Tt = enabler-team{]} $\to$ {[}Rs $>$ 5{]}}\\
  \noindent Deleted because: \texttt{({[}Tt = enabler-team{]} $\to$ {[}Au = \texttt{self-organization}{]} $\to$ {[}Rs $>$ 5{]})}

  \item \texttt{{[}Cf = $\mathsf{true}${]} $\to$ {[}Ks $>$ 5{]}}\\
  \noindent Deleted because: \texttt{({[}Cf = $\mathsf{true}${]} $\to$ {[}Aa = $\mathsf{true}${]} $\to$ {[}Ks $>$ 5{]})}

  \item \texttt{{[}Cf = $\mathsf{true}${]} $\to$ $\neg$ {[}So = $\mathsf{true}${]}}\\
  \noindent Deleted because: \texttt{({[}Cf = $\mathsf{true}${]} $\to$ {[}Aa = $\mathsf{true}${]} $\to$ $\neg$ {[}So = $\mathsf{true}${]})}

  \item \texttt{{[}Tt = \textrm{enabler-team}{]} $\to$ {[}Aa = $\mathsf{true}${]}}\\
  \noindent Deleted because: \texttt{({[}Tt = \texttt{enabler-team}{]} $\to$ {[}Cf = $\mathsf{true}${]} $\to$ {[}Aa = $\mathsf{true}${]})}
\end{enumerate}

This tool can be downloaded from \url{https://github.com/SergioGilBorras/SynT/releases/tag/V-1.4.0}, where a user's guide can also be found.

\clearpage

\section{DevOps Team Structures Theory: Operationalization}\label{app:d-operationalization}

This appendix describes the complete use case mentioned in this paper. It specifies the products generated during the preliminary conceptualisation phase (concepts and propositions) as well as during the operationalisation phase (constructs and hypotheses).

\subsection{Concepts}\label{app:d1-concepts}

\begin{itemize}
\item C1. Team. It is an artificial (abstract) concept that represents a team structure of an IT department. Specifies how the activities carried out by an organisation are divided, organised and coordinated. Possible team structures are: product teams, horizontal teams (an artificial (abstract) concept), whether bridge teams or facilitator teams, development teams and operations teams. It has attributes that explain and drive the different team structures. In our context, these attributes include the description of organizational, cultural, and technological aspects. These attributes are self-organization and autonomy, blame management, alignment of dev \& ops goals, responsibility/ownership sharing, skills/knowledge sharing, stack \& tools sharing, cross-functionality, and role definition.
\item C2. Management. It is an artificial (abstract) concept that represents the management activities (project and product management, change management, team self-organization, coordination and transfer of work between teams, as well as measuring, monitoring and feedback) to ensure that software development and maintenance are systematic, organized, and quantified.
\item C3. Culture. It represents the set of common principles, values, and best practices that impact the way that people in an organization relate to each other and make decisions and actions around application life-cycle management.
\item C4. Automation. It represents the set of activities to automate the processes of the application life-cycle management and infrastructure management where applications are deployed.
\item C5. Platform. It represents the technology to support automation. It provides interfaces for automated infrastructure management and automated application life-cycle management.
\item C6. Silo. It represents the concept of silos between teams. The silo can be organizational or cultural.
\item C7. Collaboration. From the lack or eventual collaboration to daily collaboration.
\item C8. Communication. From poor/rare communication to frequent communication.
\end{itemize}

\subsection{Propositions}\label{app:d2-propositions}

\begin{itemize}
\item P1. A team culture based on responsibility/ownership sharing enables collaboration.
\item P2. Collaboration promoting reduces organizational silos/conflicts.
\item P3. Automated application life-cycle management is associated with collaboration. Collaboration impacts automated application life-cycle management and vice versa. Automation and collaboration mutually facilitate the adoption of the other, so they are complementary.
\item P4. A team culture based on knowledge sharing enables collaboration.
\item P5. If a team is characterized by cross-functionality/skills this will increase collaboration.
\item P6. Collaboration is a property of teams in which skills take precedence over roles, i.e., the role definition/attributions code; hence, if there are already separate roles, responsibilities are very clear and collaboration is not fostered or promoted.
\item P7. A collaboration-based culture requires alignment of dev \& ops goals.
\item P8. A team culture based on metrics/visibility/feedback enables collaboration.
\item P9. Responsibility/ownership sharing is a property of cross-functionality/skills.
\item P10. Responsibility/ownership sharing reduces organizational silos/conflicts.
\item P11. Responsibility/ownership sharing is a property of organizational structures that rely on an enabler (platform) team. The existence of platform teams does not lead to a separation of responsibilities but rather they become facilitators and make ownership sharing possible, unlike dev ops (bridge) teams that become new silos with their own responsibilities (e.g., deployment, monitoring, etc.).
\item P12. Responsibility/ownership sharing is a property of team self-organization \& autonomy.
\item P13. A team culture based on responsibility/ownership sharing enables communication.
\item P14. Responsibility/ownership sharing is associated with the transfer of work between teams. If there is no shared responsibility, there is necessarily a transfer of work between development to production and operation teams (and vice versa).
\item P15. Automated infrastructure management enables responsibility/ownership sharing.
\item P16. Automated application life-cycle management enables responsibility/ownership sharing.
\item P17. Skills/knowledge sharing is a property of teams characterized by cross-functionality/skills.
\item P18. Cross-functionality/skills is a property of enabler (platform) team. 
\item P19. Cross-functionality/skills reduces organizational silos/conflicts.
\item P20. If a team is characterized by cross-functionality/skills this will increase automated application life-cycle management.
\item P21. Organizational silos/conflicts make the adoption of an automated application life-cycle management difficult.
\item P22. Metrics, visibility \& feedback enables automated application life-cycle management.
\item P23. Automated application life-cycle management enables skills/knowledge sharing.
\item P24. Enabler (platform) team enables team self-organization \& autonomy.
\item P25. Enabler (platform) team provides platform servicing.
\item P26. Automated application life-cycle management is a Platform servicing.
\item P27. Automated infrastructure management is a Platform servicing.
\item P28. Enabler (platform) teams provide automated application life-cycle management.
\end{itemize}

\subsection{Constructs}\label{app:d3-constructs}

\begin{longtable}{|p{0.24\linewidth}|p{0.7\linewidth}r|}
\caption{Construct definition: Team (DevOps case)}\label{tab:d3-construct-team}\\
    \toprule
    construct name & \multicolumn{2}{p{0.75\linewidth}|}{\textbf{Team}} \\
    \midrule
    \multirow{2}[-3]{*}{\parbox{0.24\linewidth}{Concepts}} & \multicolumn{2}{p{0.75\linewidth}|}{C1. Team} \\
    \multicolumn{1}{|l|}{from which it derives} & \multicolumn{2}{p{0.75\linewidth}|}{C6. Silo} \\
    \midrule
    operational definition & \multicolumn{2}{p{0.75\linewidth}|}{Represents a team structure of an IT department.} \\
    \midrule
    scope conditions & \multicolumn{2}{p{0.75\linewidth}|}{IT department} \\
    \midrule
    dimensionality & \multicolumn{1}{p{0.375\linewidth}|}{unidimensional} & \multicolumn{1}{p{0.375\linewidth}|}{\cellcolor[rgb]{ .906,  .902,  .902}multidimensional} \\
    \midrule
    \multicolumn{3}{|c|}{DIMENSIONS} \\
    \midrule
    \multirow{5}[2]{*}{\textbf{typeTeam}} & \multicolumn{2}{p{0.75\linewidth}|}{Possible team structures are: } \\
    \multicolumn{1}{|r|}{} & \multicolumn{2}{p{0.75\linewidth}|}{·     product teams, } \\
    \multicolumn{1}{|r|}{} & \multicolumn{2}{p{0.75\linewidth}|}{·     horizontal teams — either bridge teams and enabler teams —, } \\
    \multicolumn{1}{|r|}{} & \multicolumn{2}{p{0.75\linewidth}|}{·     development teams and} \\
    \multicolumn{1}{|r|}{} & \multicolumn{2}{p{0.75\linewidth}|}{·     operation teams.} \\
    \midrule
    derived from & \multicolumn{1}{p{0.375\linewidth}|}{\cellcolor[rgb]{ .906,  .902,  .902}data } & \multicolumn{1}{p{0.375\linewidth}|}{abductive (underrepresentation)} \\
    \midrule
    type  & \multicolumn{1}{p{0.375\linewidth}|}{\cellcolor[rgb]{ .906,  .902,  .902}scalar} & \multicolumn{1}{p{0.375\linewidth}|}{collection} \\
    \midrule
    possible values & \multicolumn{2}{p{0.75\linewidth}|}{product-team, bridge-team, enabler-team, development-team, operations-team} \\
    \midrule
    \textbf{autonomy} & \multicolumn{2}{p{0.75\linewidth}|}{A team can access to all necessary information to develop, deploy, and operate the product (self-organization) or a team depends on another team/party at some point in the product life-cycle (dependent).} \\
    \midrule
    derived from & \multicolumn{1}{p{0.375\linewidth}|}{\cellcolor[rgb]{ .906,  .902,  .902}data } & \multicolumn{1}{p{0.375\linewidth}|}{abductive (underrepresentation)} \\
    \midrule
    type  & \multicolumn{1}{p{0.375\linewidth}|}{\cellcolor[rgb]{ .906,  .902,  .902}scalar} & \multicolumn{1}{p{0.375\linewidth}|}{collection} \\
    \midrule
    possible values & \multicolumn{2}{p{0.75\linewidth}|}{dependent, self-organization} \\
    \midrule
    \textbf{blame-management} & \multicolumn{2}{p{0.75\linewidth}|}{From blameless contexts to blame contexts. In blameless contexts, the development team must not say that a given issue is a problem in the infrastructure, this is, operations team responsibility. Likewise, the operations team must not say that a failure was motivated by a problem in the application, this is, development team responsibility. The teams need to focus on solving problems, not on laying the blame on others and running away from the responsibility. Usually occurs when they do not share the responsibility.} \\
    \midrule
    derived from & \multicolumn{1}{p{0.375\linewidth}|}{\cellcolor[rgb]{ .906,  .902,  .902}data } & \multicolumn{1}{p{0.375\linewidth}|}{abductive (underrepresentation)} \\
    \midrule
    type  & \multicolumn{1}{p{0.375\linewidth}|}{\cellcolor[rgb]{ .906,  .902,  .902}scalar} & \multicolumn{1}{p{0.375\linewidth}|}{collection} \\
    \midrule
    possible values & \multicolumn{2}{p{0.75\linewidth}|}{$\mathsf{true}$, $\mathsf{false}$} \\
    \midrule
    \textbf{alignment-of-devOps-goals} & \multicolumn{2}{p{0.75\linewidth}|}{From misalignment among teams that have their own interests and goals (local optimization) to alignment with business goals and global ones (product thinking).} \\
    \midrule
    derived from & \multicolumn{1}{p{0.375\linewidth}|}{\cellcolor[rgb]{ .906,  .902,  .902}data } & \multicolumn{1}{p{0.375\linewidth}|}{abductive (underrepresentation)} \\
    \midrule
    type  & \multicolumn{1}{p{0.375\linewidth}|}{\cellcolor[rgb]{ .906,  .902,  .902}scalar} & \multicolumn{1}{p{0.375\linewidth}|}{collection} \\
    \midrule
    possible values & \multicolumn{2}{p{0.75\linewidth}|}{local-optimization, product-thinking} \\
    \midrule
    \textbf{responsibility/ownership- sharing} & \multicolumn{2}{p{0.75\linewidth}|}{From shared responsibility of the products, output artefacts (e.g., databases), and tasks (e.g., NFR shared responsibility, infrastructure management shared responsibility, monitoring shared responsibility, and incident handling shared responsibility, etc.) to separate responsibilities and tasks (each team member has different responsibilities and tasks).} \\
    \midrule
    derived from & \multicolumn{1}{p{0.375\linewidth}|}{\cellcolor[rgb]{ .906,  .902,  .902}data } & \multicolumn{1}{p{0.375\linewidth}|}{abductive (underrepresentation)} \\
    \midrule
    type  & \multicolumn{1}{p{0.375\linewidth}|}{\cellcolor[rgb]{ .906,  .902,  .902}scalar} & \multicolumn{1}{p{0.375\linewidth}|}{collection} \\
    \midrule
    possible values & \multicolumn{2}{p{0.75\linewidth}|}{Real values in the interval [0, 10], with 0 indicating null sharing and 10 meaning full sharing.} \\
    \midrule
    \textbf{skills/knowledge-sharing} & \multicolumn{2}{p{0.75\linewidth}|}{From high to low levels of knowledge sharing (e.g., developers may have knowledge about infrastructure/platform, or minimal or no awareness of what is happening on the other side of the wall, aka. wall of confusion or siloization of knowledge).} \\
    \midrule
    derived from & \multicolumn{1}{p{0.375\linewidth}|}{\cellcolor[rgb]{ .906,  .902,  .902}data } & \multicolumn{1}{p{0.375\linewidth}|}{abductive (underrepresentation)} \\
    \midrule
    type  & \multicolumn{1}{p{0.375\linewidth}|}{\cellcolor[rgb]{ .906,  .902,  .902}scalar} & \multicolumn{1}{p{0.375\linewidth}|}{collection} \\
    \midrule
    possible values & \multicolumn{2}{p{0.75\linewidth}|}{Real values in the interval [0, 10], with 0 indicating null sharing and 10 meaning full sharing} \\
    \midrule
    \textbf{stack\&tools-sharing} & \multicolumn{2}{p{0.75\linewidth}|}{From a holistic view of the tools and stack—the frontend, backend, libraries, storage, kernels, and physical machine—to non-shared stack.} \\
    \midrule
    derived from & \multicolumn{1}{p{0.375\linewidth}|}{\cellcolor[rgb]{ .906,  .902,  .902}data } & \multicolumn{1}{p{0.375\linewidth}|}{abductive (underrepresentation)} \\
    \midrule
    type  & \multicolumn{1}{p{0.375\linewidth}|}{\cellcolor[rgb]{ .906,  .902,  .902}scalar} & \multicolumn{1}{p{0.375\linewidth}|}{collection} \\
    \midrule
    possible values & \multicolumn{2}{p{0.75\linewidth}|}{Real values in the interval [0, 10], with 0 indicating non-shared stack and 10 meaning shared stack} \\
    \midrule
    \textbf{role- definition/attributions} & \multicolumn{2}{p{0.75\linewidth}|}{From "skills over roles" and T-shape/DevOps engineers (aka. full stack engineers) to well-defined and differentiated roles. When true, it indicates when a team has well-defined roles and are less likely to share responsibility. When false, it indicates that the teams do not have a well-defined definition of responsibility.} \\
    \midrule
    derived from & \multicolumn{1}{p{0.375\linewidth}|}{\cellcolor[rgb]{ .906,  .902,  .902}data } & \multicolumn{1}{p{0.375\linewidth}|}{abductive (underrepresentation)} \\
    \midrule
    type  & \multicolumn{1}{p{0.375\linewidth}|}{\cellcolor[rgb]{ .906,  .902,  .902}scalar} & \multicolumn{1}{p{0.375\linewidth}|}{collection} \\
    \midrule
    possible values & \multicolumn{2}{p{0.75\linewidth}|}{$\mathsf{true}$, $\mathsf{false}$} \\
    \midrule
    \textbf{cross-functionality/skills} & \multicolumn{2}{p{0.75\linewidth}|}{From multidisciplinary/poly-skilled teams (i.e., teams with all the necessary skills such as development, infrastructure, etc.) to teams with a lack of skills/knowledge/background.} \\
    \midrule
    derived from & \multicolumn{1}{p{0.375\linewidth}|}{\cellcolor[rgb]{ .906,  .902,  .902}data } & \multicolumn{1}{p{0.375\linewidth}|}{abductive (underrepresentation)} \\
    \midrule
    type  & \multicolumn{1}{p{0.375\linewidth}|}{\cellcolor[rgb]{ .906,  .902,  .902}scalar} & \multicolumn{1}{p{0.375\linewidth}|}{collection} \\
    \midrule
    possible values & \multicolumn{2}{p{0.75\linewidth}|}{$\mathsf{true}$, $\mathsf{false}$} \\
    \midrule
    \textbf{silo(organizational)} & \multicolumn{2}{p{0.75\linewidth}|}{From non-organizational silos/barriers (e.g., co-locating teams/departments) to siloed departments and existing organizational barriers (segregated departments; frictions, mistrust, conflicts, and disagreements among silos; silos that become bottlenecks; minimal or no awareness of what is happening on the other side of the wall).} \\
    \midrule
    derived from & \multicolumn{1}{p{0.375\linewidth}|}{\cellcolor[rgb]{ .906,  .902,  .902}data } & \multicolumn{1}{p{0.375\linewidth}|}{abductive (underrepresentation)} \\
    \midrule
    type  & \multicolumn{1}{p{0.375\linewidth}|}{\cellcolor[rgb]{ .906,  .902,  .902}scalar} & \multicolumn{1}{p{0.375\linewidth}|}{collection} \\
    \midrule
    possible values & \multicolumn{2}{p{0.75\linewidth}|}{$\mathsf{true}$, $\mathsf{false}$} \\
    \bottomrule    
\end{longtable}

\begin{longtable}{|p{0.24\linewidth}|p{0.75\linewidth}r|}
\caption{Construct definition: Management (DevOps case)}\label{tab:d3-construct-management}\\
    \toprule
    construct name & \multicolumn{2}{p{0.75\linewidth}|}{\textbf{Management}} \\
    \midrule
    \multirow{2}[-3]{*}{\parbox{0.24\linewidth}{Concepts}} & \multicolumn{2}{p{0.75\linewidth}|}{C4. Management} \\
    \multicolumn{1}{|l|}{from which it derives} & \multicolumn{2}{p{0.75\linewidth}|}{} \\
    \midrule
    operational definition & \multicolumn{2}{p{0.75\linewidth}|}{It is an artificial concept that represents the management activities (project and product management, change management, team self-organization, coordination and transfer of work between teams, as well as measuring, monitoring and feedback) to ensure that software development and maintenance are systematic, organized, and quantified.} \\
    \midrule
    scope conditions & \multicolumn{2}{p{0.75\linewidth}|}{IT organization} \\
    \midrule
    dimensionality & \multicolumn{1}{p{0.375\linewidth}|}{unidimensional} & \multicolumn{1}{p{0.375\linewidth}|}{\cellcolor[rgb]{ .906,  .902,  .902}multidimensional} \\
    \midrule
    \multicolumn{3}{|c|}{DIMENSIONS} \\
    \midrule
    \textbf{Members} & \multicolumn{2}{p{0.75\linewidth}|}{Product management (related to a product or even service model) versus traditional project management. Each product or service has a team associated with it, and that team is completely responsible for the product/service (from scoping out the functionality, to architecting it, to building it, and operating it).} \\
    \midrule
    derived from & \multicolumn{1}{p{0.375\linewidth}|}{\cellcolor[rgb]{ .906,  .902,  .902}data } & \multicolumn{1}{p{0.375\linewidth}|}{abductive (underrepresentation)} \\
    \midrule
    type  & \multicolumn{1}{p{0.375\linewidth}|}{\cellcolor[rgb]{ .906,  .902,  .902}scalar} & \multicolumn{1}{p{0.375\linewidth}|}{collection} \\
    \midrule
    possible values & \multicolumn{2}{p{0.75\linewidth}|}{product-management, project-management} \\
    \midrule
    \textbf{change-management} & \multicolumn{2}{p{0.75\linewidth}|}{From managing small and frequent changes (small batch) to large and rare changes. With a small batch size (i.e. the unit of work that passes from one stage to the next stage in a process), each batch makes it through the full life-cycle quicker, at least once a day according to Fowler definition for continuous integration. With a large batch size hampers continuous integration and delivery. “Pushing 2 weeks of commits to prod is a larger batch than pushing every commit in a continuous delivery system”.} \\
    \midrule
    derived from & \multicolumn{1}{p{0.375\linewidth}|}{\cellcolor[rgb]{ .906,  .902,  .902}data } & \multicolumn{1}{p{0.375\linewidth}|}{abductive (underrepresentation)} \\
    \midrule
    type  & \multicolumn{1}{p{0.375\linewidth}|}{\cellcolor[rgb]{ .906,  .902,  .902}scalar} & \multicolumn{1}{p{0.375\linewidth}|}{collection} \\
    \midrule
    possible values & \multicolumn{2}{p{0.75\linewidth}|}{small-frequent, large-rare} \\
    \midrule
    \textbf{metrics-visibility\&feedback} & \multicolumn{2}{p{0.75\linewidth}|}{Metrics (process, value, cost, and technical metrics) provide fast and continuous feedback from users, reduce the risk and cost of deployments, get better visibility into the delivery process itself, and manage the risks of software delivery more effectively.} \\
    \midrule
    derived from & \multicolumn{1}{p{0.375\linewidth}|}{\cellcolor[rgb]{ .906,  .902,  .902}data } & \multicolumn{1}{p{0.375\linewidth}|}{abductive (underrepresentation)} \\
    \midrule
    type  & \multicolumn{1}{p{0.375\linewidth}|}{\cellcolor[rgb]{ .906,  .902,  .902}scalar} & \multicolumn{1}{p{0.375\linewidth}|}{collection} \\
    \midrule
    possible values & \multicolumn{2}{p{0.75\linewidth}|}{$\mathsf{true}$, $\mathsf{false}$} \\
    \midrule
    \textbf{transfer-of-work-between-teams} & \multicolumn{2}{p{0.75\linewidth}|}{Transfer of work and responsibilities between teams (e.g., between development and infrastructure/operation teams). Hence, the definition of development "done" may go from committing a change to running the change in production-like environments. When there is transfer of work, depending on the high/poor trust/confidence on the other team, it may generate stress in the teams.} \\
    \midrule
    derived from & \multicolumn{1}{p{0.375\linewidth}|}{\cellcolor[rgb]{ .906,  .902,  .902}data } & \multicolumn{1}{p{0.375\linewidth}|}{abductive (underrepresentation)} \\
    \midrule
    type  & \multicolumn{1}{p{0.375\linewidth}|}{\cellcolor[rgb]{ .906,  .902,  .902}scalar} & \multicolumn{1}{p{0.375\linewidth}|}{collection} \\
    \midrule
    possible values & \multicolumn{2}{p{0.75\linewidth}|}{$\mathsf{true}$, $\mathsf{false}$} \\
    \bottomrule
\end{longtable}%

\begin{longtable}{|p{0.24\linewidth}|p{0.75\linewidth}r|}
\caption{Construct definition: Culture (DevOps case)}\label{tab:d4-universes}\\
    \toprule
    construct name & \multicolumn{2}{p{0.75\linewidth}|}{\textbf{Culture}} \\
    \midrule
    \multirow{2}[-3]{*}{\parbox{0.24\linewidth}{Concepts}} & \multicolumn{2}{p{0.75\linewidth}|}{C5. Culture} \\
    \multicolumn{1}{|l|}{from which it derives} & \multicolumn{2}{p{0.75\linewidth}|}{C6. Silo} \\
    \midrule
    operational definition & \multicolumn{2}{p{0.75\linewidth}|}{It represents the set of common principles, values, and best practices that impact the way that people in an organization relate to each other and make decisions and actions around application life-cycle management.} \\
    \midrule
    scope conditions & \multicolumn{2}{p{0.75\linewidth}|}{IT organization} \\
    \midrule
    dimensionality & \multicolumn{1}{p{0.375\linewidth}|}{unidimensional} & \multicolumn{1}{p{0.375\linewidth}|}{\cellcolor[rgb]{ .906,  .902,  .902}multidimensional} \\
    \midrule
    \multicolumn{3}{|c|}{DIMENSIONS} \\
    \midrule
    \textbf{principles} & \multicolumn{2}{p{0.75\linewidth}|}{Generic concepts as customer-centric action, create with the end in mind, end-to-end responsibility, cross-functional autonomous teams, continuous improvement, automate everything you can, among others.} \\
    \midrule
    derived from & \multicolumn{1}{p{0.375\linewidth}|}{\cellcolor[rgb]{ .906,  .902,  .902}data } & \multicolumn{1}{p{0.375\linewidth}|}{abductive (underrepresentation)} \\
    \midrule
    type  & \multicolumn{1}{p{0.375\linewidth}|}{scalar} & \multicolumn{1}{p{0.375\linewidth}|}{\cellcolor[rgb]{ .906,  .902,  .902}collection} \\
    \midrule
    possible values & \multicolumn{2}{p{0.75\linewidth}|}{customer-centric-action, end-to-end-responsibility, automate-everything} \\
    \midrule
    \textbf{values} & \multicolumn{2}{p{0.75\linewidth}|}{Cultural values as collaboration, communication, transparency, etc.} \\
    \midrule
    derived from & \multicolumn{1}{p{0.375\linewidth}|}{\cellcolor[rgb]{ .906,  .902,  .902}data } & \multicolumn{1}{p{0.375\linewidth}|}{abductive (underrepresentation)} \\
    \midrule
    type  & \multicolumn{1}{p{0.375\linewidth}|}{scalar} & \multicolumn{1}{p{0.375\linewidth}|}{\cellcolor[rgb]{ .906,  .902,  .902}collection} \\
    \midrule
    possible values & \multicolumn{2}{p{0.75\linewidth}|}{(communication, collaboration, transparency)} \\
    \midrule
    \textbf{best-practices} & \multicolumn{2}{p{0.75\linewidth}|}{DevOps best/good practices} \\
    \midrule
    derived from & \multicolumn{1}{p{0.375\linewidth}|}{\cellcolor[rgb]{ .906,  .902,  .902}data } & \multicolumn{1}{p{0.375\linewidth}|}{abductive (underrepresentation)} \\
    \midrule
    type  & \multicolumn{1}{p{0.375\linewidth}|}{scalar} & \multicolumn{1}{p{0.375\linewidth}|}{\cellcolor[rgb]{ .906,  .902,  .902}collection} \\
    \midrule
    possible values & \multicolumn{2}{p{0.75\linewidth}|}{(Continuous-integration, continuous-testing, continuous-delivery, continuous-deployment, continuous-monitoring, continuous-improvement, infrastructure-as-code)} \\
    \midrule
    \textbf{silo(cultural)} & \multicolumn{2}{p{0.75\linewidth}|}{From non-cultural barriers to existing cultural barriers. Not to be confused with organizational/silos conflicts. While the cultural silos/conflicts focus on practices and culture that leads/break to barriers, organization silos/conflicts focus on team structures.} \\
    \midrule
    derived from & \multicolumn{1}{p{0.375\linewidth}|}{\cellcolor[rgb]{ .906,  .902,  .902}data } & \multicolumn{1}{p{0.375\linewidth}|}{abductive (underrepresentation)} \\
    \midrule
    type  & \multicolumn{1}{p{0.375\linewidth}|}{\cellcolor[rgb]{ .906,  .902,  .902}scalar} & \multicolumn{1}{p{0.375\linewidth}|}{collection} \\
    \midrule
    possible values & \multicolumn{2}{p{0.75\linewidth}|}{$\mathsf{true}$, $\mathsf{false}$} \\
    \bottomrule
\end{longtable}%

\begin{longtable}{|p{0.24\linewidth}|p{0.75\linewidth}r|}
\caption{Construct definition: Automation (DevOps case)}\label{tab:d3-construct-automation}\\
    \toprule
    construct name & \multicolumn{2}{p{0.75\linewidth}|}{\textbf{Automation}} \\
    \midrule
    \multirow{2}[-3]{*}{\parbox{0.24\linewidth}{Concepts}} & \multicolumn{2}{p{0.75\linewidth}|}{C6. Automation} \\
    \multicolumn{1}{|l|}{from which it derives} & \multicolumn{2}{p{0.75\linewidth}|}{C7. Platform} \\
    \midrule
    operational definition & \multicolumn{2}{p{0.75\linewidth}|}{Represents the set of activities to automate the processes of the application lifecycle management and infrastructure management where applications are deployed.} \\
    \midrule
    scope conditions & \multicolumn{2}{p{0.75\linewidth}|}{IT department} \\
    \midrule
    dimensionality & \multicolumn{1}{p{0.375\linewidth}|}{unidimensional} & \multicolumn{1}{p{0.375\linewidth}|}{\cellcolor[rgb]{ .906,  .902,  .902}multidimensional} \\
    \midrule
    \multicolumn{3}{|c|}{DIMENSIONS} \\
    \midrule
    \textbf{platform} & \multicolumn{2}{p{0.75\linewidth}|}{It represents the technology to support automation. It provides interfaces for automated infrastructure management and automated application life-cycle management.} \\
    \midrule
    derived from & \multicolumn{1}{p{0.375\linewidth}|}{\cellcolor[rgb]{ .906,  .902,  .902}data } & \multicolumn{1}{p{0.375\linewidth}|}{abductive (underrepresentation)} \\
    \midrule
    type  & \multicolumn{1}{p{0.375\linewidth}|}{\cellcolor[rgb]{ .906,  .902,  .902}scalar} & \multicolumn{1}{p{0.375\linewidth}|}{collection} \\
    \midrule
    possible values & \multicolumn{2}{p{0.75\linewidth}|}{automated-infrastructure-management-interface, automated-application-life-cycle-management-interface} \\
    \midrule
    \textbf{automated-application- life-cycle-management} & \multicolumn{2}{p{0.75\linewidth}|}{Represents pipelines and tools for automated delivery/deployment and operation.} \\
    \midrule
    derived from & \multicolumn{1}{p{0.375\linewidth}|}{\cellcolor[rgb]{ .906,  .902,  .902}data } & \multicolumn{1}{p{0.375\linewidth}|}{abductive (underrepresentation)} \\
    \midrule
    type  & \multicolumn{1}{p{0.375\linewidth}|}{\cellcolor[rgb]{ .906,  .902,  .902}scalar} & \multicolumn{1}{p{0.375\linewidth}|}{collection} \\
    \midrule
    possible values & \multicolumn{2}{p{0.75\linewidth}|}{$\mathsf{true}$, $\mathsf{false}$} \\
    \midrule
    \textbf{automated-infrastructure-management} & \multicolumn{2}{p{0.75\linewidth}|}{Represents techniques, technologies, and tools for automated infrastructure management.} \\
    \midrule
    derived from & \multicolumn{1}{p{0.375\linewidth}|}{\cellcolor[rgb]{ .906,  .902,  .902}data } & \multicolumn{1}{p{0.375\linewidth}|}{abductive (underrepresentation)} \\
    \midrule
    type  & \multicolumn{1}{p{0.375\linewidth}|}{\cellcolor[rgb]{ .906,  .902,  .902}scalar} & \multicolumn{1}{p{0.375\linewidth}|}{collection} \\
    \midrule
    possible values & \multicolumn{2}{p{0.75\linewidth}|}{$\mathsf{true}$, $\mathsf{false}$} \\
    \midrule
    \textbf{platform-servicing} & \multicolumn{2}{p{0.75\linewidth}|}{Services that an entity (e.g., a team, an external consultant, etc.) offers/provides to product teams (developers and/or operators) to assist them on DevOps platform, such as: infrastructure management (e.g., containerization, cloud, etc.), infrastructure automation (from low to high levels of automated infrastructure services aka. Infrastructure as Code, IaC), pipeline automation (CI/CD and release tools), IT operation, etc. This set of services is supported by Enabler-team.} \\
    \midrule
    derived from & \multicolumn{1}{p{0.375\linewidth}|}{\cellcolor[rgb]{ .906,  .902,  .902}data } & \multicolumn{1}{p{0.375\linewidth}|}{abductive (underrepresentation)} \\
    \midrule
    type  & \multicolumn{1}{p{0.375\linewidth}|}{\cellcolor[rgb]{ .906,  .902,  .902}scalar} & \multicolumn{1}{p{0.375\linewidth}|}{collection} \\
    \midrule
    possible values & \multicolumn{2}{p{0.75\linewidth}|}{$\mathsf{true}$, $\mathsf{false}$} \\
    \bottomrule
\end{longtable}%

\begin{longtable}{|p{0.24\linewidth}|p{0.75\linewidth}r|}
\caption{Construct definition: Interaction (DevOps case)}\label{tab:d3-construct-interaction}\\
    \toprule
    construct name & \multicolumn{2}{p{0.75\linewidth}|}{\textbf{Interaction}} \\
    \midrule
    \multirow{2}[-3]{*}{\parbox{0.24\linewidth}{Concepts}} & \multicolumn{2}{p{0.75\linewidth}|}{C7. Collaboration} \\
    \multicolumn{1}{|l|}{from which it derives} & \multicolumn{2}{p{0.375\linewidth}|}{C8. Communication} \\
    \midrule
    operational definition & \multicolumn{2}{p{0.75\linewidth}|}{Represents interactions in a team measured as collaborations and communications.} \\
    \midrule
    scope conditions & \multicolumn{2}{p{0.75\linewidth}|}{teams} \\
    \midrule
    dimensionality & \multicolumn{1}{p{0.375\linewidth}|}{unidimensional} & \multicolumn{1}{p{0.375\linewidth}|}{\cellcolor[rgb]{ .906,  .902,  .902}multidimensional} \\
    \midrule
    \multicolumn{3}{|c|}{DIMENSIONS} \\
    \midrule
    \textbf{collaboration} & \multicolumn{2}{p{0.75\linewidth}|}{The frequency and quality of collaboration between teams when they share responsibility for an activity. From eventual collaboration, which may generate conflicts and disagreements on decisions (low quality), to daily collaboration (working together regularly on a daily basis) with high quality.} \\
    \midrule
    derived from & \multicolumn{1}{p{0.375\linewidth}|}{\cellcolor[rgb]{ .906,  .902,  .902}data } & \multicolumn{1}{p{0.375\linewidth}|}{\cellcolor[rgb]{ .906,  .902,  .902}abductive (underrepresentation)} \\
    \midrule
    type  & \multicolumn{1}{p{0.375\linewidth}|}{\cellcolor[rgb]{ .906,  .902,  .902}scalar} & \multicolumn{1}{p{0.375\linewidth}|}{collection} \\
    \midrule
    possible values & \multicolumn{2}{p{0.75\linewidth}|}{(Daily, High), (Daily, Low), (Eventual, High), (Eventual, Low)} \\
    \midrule
    \textbf{communication} & \multicolumn{2}{p{0.75\linewidth}|}{Represents the level of communication established between the teams. From poor/rare communication (and standardization) to frequent communication.} \\
    \midrule
    derived from & \multicolumn{1}{p{0.375\linewidth}|}{\cellcolor[rgb]{ .906,  .902,  .902}data } & \multicolumn{1}{p{0.375\linewidth}|}{abductive (underrepresentation)} \\
    \midrule
    type  & \multicolumn{1}{p{0.375\linewidth}|}{\cellcolor[rgb]{ .906,  .902,  .902}scalar} & \multicolumn{1}{p{0.375\linewidth}|}{collection} \\
    \midrule
    possible values & \multicolumn{2}{p{0.75\linewidth}|}{$\mathsf{true}$, $\mathsf{false}$} \\
    \bottomrule
\end{longtable}%

\subsection{Universes}\label{app:d4-universes}

\begin{longtable}{|r|p{0.75\linewidth}|}
   \caption{Universes (types), admissible values, and variables for the DevOps case study}\label{tab:universes}\\
    \toprule
    \multicolumn{2}{|c|}{\textbf{Tty}} \\
    \midrule
    \multicolumn{1}{|p{0.24\linewidth}|}{values} & product-team, bridge-team, enabler-team, development-team, operations-team \\
    \midrule
    \multicolumn{1}{|p{0.24\linewidth}|}{variables with this type} & typeTeam (Tt) (Team) \\
    \midrule
    \multicolumn{1}{|p{0.24\linewidth}|}{functions} & None \\
    \midrule
    \multicolumn{1}{|p{0.24\linewidth}|}{relations} & = \\
    \midrule
    \multicolumn{2}{|c|}{\textbf{Tau}} \\
    \midrule
    \multicolumn{1}{|p{0.24\linewidth}|}{values} & dependent, self-organization \\
    \midrule
    \multicolumn{1}{|p{0.24\linewidth}|}{variables with this type} & autonomy (Au) (Team) \\
    \midrule
    \multicolumn{1}{|p{0.24\linewidth}|}{functions} & None \\
    \midrule
    \multicolumn{1}{|p{0.24\linewidth}|}{relations (arity)} & $=$, $>$, where self-organization $>$ dependent \\
    \midrule
    \multicolumn{2}{|c|}{\textbf{Tag}} \\
    \midrule
    \multicolumn{1}{|p{0.24\linewidth}|}{values} & local-optimization, product-thinking \\
    \midrule
    \multicolumn{1}{|p{0.24\linewidth}|}{variables with this type} & alignment-of-devOps-goals (Ag) (Team) \\
    \midrule
    \multicolumn{1}{|p{0.24\linewidth}|}{functions} & None \\
    \midrule
    \multicolumn{1}{|p{0.24\linewidth}|}{relations} & $=$, $>$, where product-thinking $>$ local-optimization \\
    \midrule
    \multicolumn{2}{|c|}{\textbf{Tre}} \\
    \midrule
    \multicolumn{1}{|p{0.24\linewidth}|}{values} & Interval [0,10] \\
    \midrule
    \multicolumn{1}{|p{0.24\linewidth}|}{variables with this type} & responsibility/ownership- sharing (Rs) (Team), skills/knowledge-sharing (Ks) (Team), stack\&tools-sharing (Ss) (Team) \\
    \midrule
    \multicolumn{1}{|p{0.24\linewidth}|}{functions} & None \\
    \midrule
    \multicolumn{1}{|p{0.24\linewidth}|}{relations} & $=$, $>$ \\
    \midrule
    \multicolumn{2}{|c|}{\textbf{Tbo}} \\
    \midrule
    \multicolumn{1}{|p{0.24\linewidth}|}{values} & $\mathsf{true}$, $\mathsf{false}$ \\
    \midrule
    \multicolumn{1}{|p{0.24\linewidth}|}{variables with this type} & blame-management (Bm) (Team), role-definition/attributions (Rd) (Team), cross-functionality/skills (Cf) (Team), silo(organizational) (So) (Team), metrics-visibility\&feedback (Mv) (Management), transfer-of-work-between-teams (Tw (Management)), silo(cultural) (Sc) (Culture), automated-application-life-cycle-management (Aa) (Automation), automated-infrastructure-management (Ai) (Automation), platform-servicing (Ps) (Automation), communication (Cm) (Interaction) \\
    \midrule
    \multicolumn{1}{|p{0.24\linewidth}|}{functions} & None \\
    \midrule
    \multicolumn{1}{|p{0.24\linewidth}|}{relations} & = \\
    \midrule
    \multicolumn{2}{|c|}{\textbf{Tma}} \\
    \midrule
    \multicolumn{1}{|p{0.24\linewidth}|}{values} & product-management, project-management \\
    \midrule
    \multicolumn{1}{|p{0.24\linewidth}|}{variables with this type} & members (Me) (Management) \\
    \midrule
    \multicolumn{1}{|p{0.24\linewidth}|}{functions} & None \\
    \midrule
    \multicolumn{1}{|p{0.24\linewidth}|}{relations} & = \\
    \midrule
    \multicolumn{2}{|c|}{\textbf{Tcm}} \\
    \midrule
    \multicolumn{1}{|p{0.24\linewidth}|}{values} & small-frequent, large-rare \\
    \midrule
    \multicolumn{1}{|p{0.24\linewidth}|}{variables with this type} & change-management (Ch) (Management) \\
    \midrule
    \multicolumn{1}{|p{0.24\linewidth}|}{functions} & None \\
    \midrule
    \multicolumn{1}{|p{0.24\linewidth}|}{relations} & =, $>$, where small-frequent $>$ large-rare \\
    \midrule
    \multicolumn{2}{|c|}{\textbf{Tpr}} \\
    \midrule
    \multicolumn{1}{|p{0.24\linewidth}|}{values} & collection (customer-centric-action, end-to-end-responsibility, automate-everything) \\
    \midrule
    \multicolumn{1}{|p{0.24\linewidth}|}{variables with this type} & principles (Pr) (Culture) \\
    \midrule
    \multicolumn{1}{|p{0.24\linewidth}|}{functions} & None \\
    \midrule
    \multicolumn{1}{|p{0.24\linewidth}|}{relations} & = \\
    \midrule
    \multicolumn{2}{|c|}{\textbf{Tva}} \\
    \midrule
    \multicolumn{1}{|p{0.24\linewidth}|}{values} & collection (communication, collaboration, transparency) \\
    \midrule
    \multicolumn{1}{|p{0.24\linewidth}|}{variables with this type} & values (Va) (Culture) \\
    \midrule
    \multicolumn{1}{|p{0.24\linewidth}|}{functions} & None \\
    \midrule
    \multicolumn{1}{|p{0.24\linewidth}|}{relations} & = \\
    \midrule
   \multicolumn{2}{|c|}{\textbf{Tbp}} \\
    \midrule
    \multicolumn{1}{|p{0.24\linewidth}|}{values} & collection (Continuous-integration, continuous-testing, continuous-delivery, continuous-deployment, continuous-monitoring, continuous-improvement, infrastructure-as-code) \\
    \midrule
    \multicolumn{1}{|p{0.24\linewidth}|}{variables with this type} & best-practices (Bp) (Culture) \\
    \midrule
    \multicolumn{1}{|p{0.24\linewidth}|}{functions} & None \\
    \midrule
    \multicolumn{1}{|p{0.24\linewidth}|}{relations} & = \\
    \midrule
    \multicolumn{2}{|c|}{\textbf{Tat}} \\
    \midrule
    \multicolumn{1}{|p{0.24\linewidth}|}{values} & automated-infrastructure-management-interface, automated-application-life-cycle-management-interface \\
    \midrule
    \multicolumn{1}{|p{0.24\linewidth}|}{variables with this type} & platform (Pl) (Automation) \\
    \midrule
    \multicolumn{1}{|p{0.24\linewidth}|}{functions} & None \\
    \midrule
    \multicolumn{1}{|p{0.24\linewidth}|}{relations} & = \\
    \midrule
    \multicolumn{2}{|c|}{\textbf{Tco}} \\
    \midrule
    \multicolumn{1}{|p{0.24\linewidth}|}{values} & (Daily, High), (Eventual, Low), (Daily, Low), (Eventual, High) \\
    \midrule
    \multicolumn{1}{|p{0.24\linewidth}|}{variables with this type} & collaboration (Co) (Interaction) \\
    \midrule
    \multicolumn{1}{|p{0.24\linewidth}|}{functions} & None \\
    \midrule
    \multicolumn{1}{|l|}{\multirow{6}[2]{*}{relations}} & $=$, $>$, where: \\
          & (Daily, High) $>$ (Daily, Low), \\
          & (Daily, High) $>$ (Eventual, High), \\
          & (Daily, Low) $>$ (Eventual, Low), \\
          & (Eventual, High) $>$ (Eventual, Low), \\
          & (Daily, High) $>$ (Eventual, Low) \\
    \bottomrule
\end{longtable}%

\newpage

Therefore, associated language $\mathcal{L} = \{\mathcal{U}, \mathcal{V}, \mathcal{F}, \mathcal{R}\},$ is specified as
$$
\mathcal{U} = (Tty, Tau, Tag, Tre, Tbo, Tma, Tcm, Tpr, Tva, Tbp, Tat, Tco),
$$
where:
\begingroup
\begin{align*}
Tty &= \{\text{product-team}, \text{bridge-team}, \text{enabler-team}, \text{development-team}, \text{operations-team}\},\\
Tau &= \{\text{dependent}, \text{self-organization}\},\\
Tag &= \{\text{local-optimization}, \text{product-thinking}\},\\
Tre &= \{\text{Interval }[0,10]\},\\
Tbo &= \{\mathsf{true}, \mathsf{false}\},\\
Tma &= \{\text{product-management}, \text{project-management}\},\\
Tcm &= \{\text{small-frequent}, \text{large-rare}\},\\
Tpr &= \{\text{collection }(\text{customer-centric-action}, \text{end-to-end-responsibility}, \text{automate-everything})\},\\
Tva &= \{\text{collection }(\text{communication}, \text{collaboration}, \text{transparency})\},\\
Tbp &= \{\text{collection }(\text{Continuous-}\allowbreak\text{integration},\allowbreak
\text{continuous-testing},\allowbreak \text{continuous-delivery},\allowbreak \text{continuous-deployment},\\
&\qquad \text{continuous-monitoring},\allowbreak \text{continuous-improvement},\allowbreak
\text{infrastructure-}\allowbreak\text{as-}\allowbreak\text{code})\},\\
Tat &= \{\text{automated-}\allowbreak\text{infrastructure-}\allowbreak\text{management-interface},\allowbreak
\text{automated-}\allowbreak\text{application-}\allowbreak\text{life-cycle-}\allowbreak\text{management-interface}\},\\
Tco &= \{\text{(Daily, High)},\allowbreak \text{(Eventual, Low)},\allowbreak \text{(Daily, Low)},\allowbreak \text{(Eventual, High)}\}.
\end{align*}
\endgroup

For the variables, functions and relations, we have
\begin{align*}
  \mathcal{V} &= (\{Tt\}, \{Au\}, \{Ag\}, \{Rs, Ks, Ss\}, \{Bm, Rd, Cf, So, Mv, Tw, Sc, Aa, Ai, Ps, Cm\},\\
&\qquad \{Me\}, \{Ch\}, \{Pr\}, \{Va\}, \{Bp\}, \{Pl\}, \{Co\}),\\
\mathcal{F} &= (\varnothing, \varnothing, \varnothing, \varnothing, \varnothing, \varnothing,\varnothing, \varnothing, \varnothing, \varnothing, \varnothing, \varnothing),\\
\mathcal{R} &= (\varnothing, \{>\}, \{>\}, \{>\}, \{=\}, \varnothing, \{>\}, \varnothing, \varnothing, \varnothing, \varnothing, \{=, >\}).
\end{align*}

\subsection{Hypotheses}\label{app:d5-hypotheses}

\begin{itemize}
  \item P1. ``\emph{A team culture based on responsibility/ownership sharing
enables collaboration}''. If we consider that a company has achieved a
culture based on ownership sharing when Rs $>$ 5, and that
collaboration exists when we have more than (Eventual, Low)
collaboration, then the resulting formula is:

\begin{align*}
\varphi_{1} &= {[}Rs > 5{]} \to {[}Co > \textrm{(Eventual, Low)}{]},\\
\varphi_{1}' &= \neg {[}Co > \textrm{(Eventual, Low)}{]} \to \neg {[}Rs > 5{]}.
\end{align*}

As the tool that has been developed automatically calculates and enters counter-reciprocal theory, we will not describe it in this section. What is interesting when entering data into the tool is to indicate the literals of each of these hypotheses. In the case at hand, they are:

\begin{align*}
l_1 &= {[}Rs > 5{]},\\
l_2 &= {[}Co > \textrm{(Eventual, Low)}{]}.
\end{align*}

As the tool automatically calculates and enters the negated form of the literals when entering the counter-reciprocal, we will not detail these negated literals here.	

\item P2. ``\emph{Promoting collaboration is associated with lower organizational
silos/conflicts}''. With the same interpretation that collaboration
exists when it holds $Co > \textrm{(Eventual, Low)}$, we obtain:
$$\varphi_{2} = {[}Co > \textrm{(Eventual, Low)}{]} \to \neg {[}So = \mathsf{true}{]}$$
$$l_2 = {[}Co > \textrm{(Eventual, Low)}{]}\qquad l_3 = {[}So = \mathsf{true}{]}$$

\item P3. ``\emph{Automated application life-cycle management is associated
with collaboration. Collaboration impacts automated application
life-cycle management and vice versa. Automation and collaboration
mutually facilitate the adoption of the other, so they are
complementary}''. This proposition establishes a double implication that we express with the following two formulas:
$$\varphi_{3} = {[}Co > \textrm{(Eventual, Low)}{]} \to {[}Aa = \mathsf{true}{]}$$
$$\varphi_{4} = {[}Aa = \mathsf{true}{]} \to {[}Co > \textrm{(Eventual, Low)}{]}$$
$$l_2 = {[}Co > \textrm{(Eventual, Low)}{]}\qquad l_4 = {[}Aa = \mathsf{true}{]}$$

\item P4. ``\emph{A team culture based on knowledge sharing enables
collaboration}''.
$$\varphi_{5} = {[}Ks > 5{]} \to {[}Co > \textrm{(Eventual, Low)}{]}$$
$$l_2 = {[}Co > \textrm{(Eventual, Low)}{]}\qquad l_5 = {[}Ks > 5{]}$$

\item P5. ``\emph{If a team is characterized by cross-functionality/skills,
this is associated with higher collaboration}''. In this case, ``enables collaboration'' has the same meaning as ``increase collaboration''. Thus, this proposition is expressed by the following formula:
$$\varphi_{6} = {[}Cf = \mathsf{true}{]} \to {[}Co > \textrm{(Eventual, Low)}{]}$$
$$l_2 = {[}Co > \textrm{(Eventual, Low)}{]}\qquad l_6 = {[}Cf = \mathsf{true}{]}$$

\item P6. ``\emph{Collaboration is a property of teams in which skills take
precedence over roles, i.e., the role definition/attributions code;
hence, if there are already separate roles, responsibilities are very
clear and collaboration is not fostered or promoted}''. Here the consequent is negated (no collaboration is expected), yielding:
$$\varphi_{7} = {[}Rd = \mathsf{true}{]} \to \neg {[}Co > \textrm{(Eventual, Low)}{]}$$
$$l_2 = {[}Co > \textrm{(Eventual, Low)}{]}\qquad l_7 = {[}Rd = \mathsf{true}{]}$$

\item P7. ``\emph{A collaboration-based culture requires alignment of dev \&
ops goals}''. In this case, alignment with the objectives is considered
to exist if {[}$Ag > \text{local-optimization}${]} according to the
relationship established in Appendix~\ref{app:d4-universes}, so we obtain:
$$\varphi_{8} = {[}Co > \textrm{(Eventual, Low)}{]} \to {[}Ag > \text{local-optimization}{]}$$
$$l_2 = {[}Co > \textrm{(Eventual, Low)}{]}\qquad l_8 = {[}Ag > \text{local-optimization}{]}$$

\item P8. ``\emph{A team culture based on metrics/visibility/feedback enables
collaboration}''. Regularly performing measurement and sharing
activities is associated with a more collaborative culture.
$$\varphi_{9} = {[}Mv = \mathsf{true}{]} \to {[}Co > \textrm{(Eventual, Low)}{]}$$
$$l_2 = {[}Co > \textrm{(Eventual, Low)}{]}\qquad l_9 = {[}Mv = \mathsf{true}{]}$$

\item P9. ``\emph{Responsibility/ownership sharing is a property of
cross-functionality/skills}''. The cross-functionality/skills teams are
characterised by sharing all or some responsibility of products, output
artefacts (e.g., databases), and tasks (e.g., NFR shared responsibility,
infrastructure management shared responsibility, monitoring shared
responsibility, and incident handling shared responsibility, etc.) so
that everyone is responsible for build, test, deploy, operate and
maintain their applications and infrastructure (although some
responsibilities may still reside on one side -dev or ops-).
$$\varphi_{10} = {[}Cf = \mathsf{true}{]} \to {[}Rs > 5{]}$$
$$l_1 = {[}Rs > 5{]}\qquad l_6 = {[}Cf = \mathsf{true}{]}$$

\item P10. ``\emph{Responsibility/ownership sharing is associated with lower organizational
silos/conflicts}''.
$$\varphi_{11} = {[}Rs > 5{]} \to \neg {[}So = \mathsf{true}{]}$$
$$l_1 = {[}Rs > 5{]}\qquad l_3 = {[}So = \mathsf{true}{]}$$

\item P11. ``\emph{Responsibility/ownership sharing is a property of
organizational structures that rely on an enabler (platform) team. The
existence of platform teams does not lead to a separation of
responsibilities but rather they become facilitators and make ownership
sharing possible, unlike DevOps (bridge) teams that become new silos
with their own responsibilities (e.g., deployment, monitoring, etc.)}''.
This proposition affects the variable Tt and two of the values it can
take: enabler-team and bridge-team, so we will use two formulas to
represent it:
$$\varphi_{12} = {[}Tt = \textrm{enabler-team}{]} \to {[}Rs > 5{]}$$
$$\varphi_{13} = {[}Tt = \textrm{bridge-team}{]} \to {[}So = \mathsf{true}{]}$$
$$l_1 = {[}Rs > 5{]}\qquad l_3 = {[}So = \mathsf{true}{]}$$
$$l_{10} = {[}Tt = \textrm{enabler-team}{]}\qquad l_{11} = {[}Tt = \textrm{bridge-team}{]}$$

\item P12. ``\emph{Responsibility/ownership sharing is a property of team
self-organization \& autonomy}''. If a team in the autonomy attribute
has the value self-organisation, this implies that there is
responsibility/ownership sharing.
$$\varphi_{14} = {[}Au = \textrm{self-organization}{]} \to {[}Rs > 5{]}$$
$$l_1 = {[}Rs > 5{]}\qquad l_{12} = {[}Au = \textrm{self-organization}{]}$$

\item P13. ``\emph{A team culture based on responsibility/ownership sharing
enables communication}''.
$$\varphi_{15} = {[}Rs > 5{]} \to {[}Cm = \mathsf{true}{]}$$
$$l_1 = {[}Rs > 5{]}\qquad l_{13} = {[}Cm = \mathsf{true}{]}$$

\item P14. ``\emph{Responsibility/ownership sharing is associated with the
transfer of work between teams. If there is no shared responsibility,
there is necessarily a transfer of work between development and
production/operations teams (and vice versa)}''.
$$\varphi_{16} = \neg {[}Rs > 5{]} \to {[}Tw = \mathsf{true}{]}$$
$$\varphi_{17} = {[}Tw = \mathsf{true}{]} \to \neg {[}Rs > 5{]}$$
$$l_1 = {[}Rs > 5{]}\qquad l_{14} = {[}Tw = \mathsf{true}{]}$$

\item P15. ``\emph{Automated infrastructure management enables
responsibility/ownership sharing}''.
$$\varphi_{18} = {[}Ai = \mathsf{true}{]} \to {[}Rs > 5{]}$$
$$l_1 = {[}Rs > 5{]}\qquad l_{15} = {[}Ai = \mathsf{true}{]}$$

\item P16. ``\emph{Automated application life-cycle management enables
responsibility/ownership sharing}''.
$$\varphi_{19} = {[}Aa = \mathsf{true}{]} \to {[}Rs > 5{]}$$
$$l_1 = {[}Rs > 5{]}\qquad l_4 = {[}Aa = \mathsf{true}{]}$$

\item P17. ``\emph{Skills/knowledge sharing is a property of teams
characterized by cross-functionality/skills}''.
$$\varphi_{20} = {[}Cf = \mathsf{true}{]} \to {[}Ks > 5{]}$$
$$l_5 = {[}Ks > 5{]}\qquad l_6 = {[}Cf = \mathsf{true}{]}$$

\item P18. ``\emph{Cross-functionality/skills is a property of enabler
(platform) team}''.
$$\varphi_{21} = {[}Tt = \textrm{enabler-team}{]} \to {[}Cf = \mathsf{true}{]}$$
$$l_6 = {[}Cf = \mathsf{true}{]}\qquad l_{10} = {[}Tt = \textrm{enabler-team}{]}$$

\item P19. ``\emph{Cross-functionality/skills is associated with lower organizational
silos/conflicts}''. Multidisciplinary/poly-skilled teams (i.e., teams
with all the necessary skills such as development, infrastructure, etc.)
avoid organisational silos.
$$\varphi_{22} = {[}Cf = \mathsf{true}{]} \to \neg {[}So = \mathsf{true}{]}$$
$$l_3 = {[}So = \mathsf{true}{]}\qquad l_6 = {[}Cf = \mathsf{true}{]}$$

\item P20. ``\emph{If a team is characterized by cross-functionality/skills
this is associated with higher adoption of automated application life-cycle management}''.
Multidisciplinary/poly-skilled teams will increase automated application life-cycle management.
$$\varphi_{23} = {[}Cf = \mathsf{true}{]} \to {[}Aa = \mathsf{true}{]}$$
$$l_4 = {[}Aa = \mathsf{true}{]}\qquad l_6 = {[}Cf = \mathsf{true}{]}$$

\item P21. ``\emph{Organizational silos/conflicts are associated with lower adoption of
automated application life-cycle management}''.
$$\varphi_{24} = {[}So = \mathsf{true}{]} \to \neg {[}Aa = \mathsf{true}{]}$$
$$l_3 = {[}So = \mathsf{true}{]}\qquad l_4 = {[}Aa = \mathsf{true}{]}$$

\item P22. ``\emph{Metrics, visibility \& feedback enables automated
application life-cycle management}''. Metrics (process, value, cost, and technical metrics) provide fast and continuous feedback from users, reduce the risk and cost of deployments, get better visibility into the delivery process itself, and manage the risks of software delivery more effectively.
$$\varphi_{25} = {[}Mv = \mathsf{true}{]} \to {[}Aa = \mathsf{true}{]}$$
$$l_4 = {[}Aa = \mathsf{true}{]}\qquad l_9 = {[}Mv = \mathsf{true}{]}$$

\item P23. ``\emph{Automated application life-cycle management enables
skills/knowledge sharing}''.
$$\varphi_{26} = {[}Aa = \mathsf{true}{]} \to {[}Ks > 5{]}$$
$$l_4 = {[}Aa = \mathsf{true}{]}\qquad l_5 = {[}Ks > 5{]}$$

\item P24. ``\emph{Enabler (platform) team enables team self-organization \&
autonomy}''. An enabling team is characterised by
cross-functionality/skills and is associated with autonomy:
self-organization (can access all the information needed to develop,
deploy and operate the product) without having to wait for anyone else.
$$\varphi_{27} = {[}Tt = \textrm{enabler-team}{]} \to {[}Au = \text{\textrm{self-organization}}{]}$$
$$l_{10} = {[}Tt = \textrm{enabler-team}{]}\qquad l_{12} = {[}Au = \text{\textrm{self-organization}}{]}$$

\item P25. ``\emph{Enabler (platform) team provides platform servicing}''.
Enabler team provides platform servicing to product teams to assist them
on the DevOps platform.
$$\varphi_{28} = {[}Tt = \textrm{enabler-team}{]} \to {[}Ps = \mathsf{true}{]}$$
$$l_{10} = {[}Tt = \textrm{enabler-team}{]}\qquad l_{16} = {[}Ps = \mathsf{true}{]}$$

\item P26. Automated application life-cycle management is platform servicing.

\item P27. Automated infrastructure management is platform servicing.

P26 and P27 are considered atomic propositions and thus do not generate any
hypotheses.

\item P28. ``\emph{Enabler (platform) teams provide automated application
life-cycle management}''. Enabler platform teams provide automated application life-cycle management which is nothing more than a special case of platform servicing.
$$\varphi_{29} = {[}Tt = \textrm{enabler-team}{]} \to {[}Aa = \mathsf{true}{]}$$
$$l_4 = {[}Aa = \mathsf{true}{]}\qquad l_{10} = {[}Tt = \textrm{enabler-team}{]}$$
\end{itemize}

\end{document}